\documentclass{aa}  
\usepackage{xcolor}
\usepackage{layouts}
\usepackage{graphicx}
\usepackage{todonotes}
\usepackage{rotating} 
\usepackage{caption}

\newcommand{\hb}{H$_\beta$ }
\newcommand{\ha}{H$_\alpha$ }
\newcommand{\hg}{H$_\gamma$ }

\def\deg{\mbox{$^{\circ}$}}

\usepackage[breaklinks=true]{hyperref}
\usepackage{natbib,twoopt}
\bibpunct{(}{)}{;}{a}{}{,}             
\makeatletter
  \newcommandtwoopt{\citeads}[3][][]{\href{http://adsabs.harvard.edu/abs/#3}%
    {\def\hyper@linkstart##1##2{}%
     \let\hyper@linkend\@empty\citealp[#1][#2]{#3}}}
  \newcommandtwoopt{\citepads}[3][][]{\href{http://adsabs.harvard.edu/abs/#3}%
    {\def\hyper@linkstart##1##2{}%
     \let\hyper@linkend\@empty\citep[#1][#2]{#3}}}
  \newcommandtwoopt{\citetads}[3][][]{\href{http://adsabs.harvard.edu/abs/#3}%
    {\def\hyper@linkstart##1##2{}%
     \let\hyper@linkend\@empty\citet[#1][#2]{#3}}}
  \newcommandtwoopt{\citeyearads}[3][][]%
    {\href{http://adsabs.harvard.edu/abs/#3}
    {\def\hyper@linkstart##1##2{}%
     \let\hyper@linkend\@empty\citeyear[#1][#2]{#3}}}
\makeatother
\usepackage{graphicx}
\usepackage{booktabs}
\usepackage{lipsum}
\usepackage{txfonts}
\usepackage{hyperref}
\usepackage{tabularx}
\usepackage{upgreek}

\graphicspath{{Figures/}} 

\begin{document} 
\abstract
   {The nuclear X-ray transient eRASSt J012026.5-292727 (J012026 hereafter) was discovered in the second SRG/eROSITA all-sky survey (eRASS2). The source appeared more than one order of magnitude brighter than the eRASS1 upper limits (peak eRASS2 0.2--2.3~keV flux of $1.14 \times 10^{-12}$~erg~cm$^{-2}$~s$^{-1}$), and with a soft X-ray spectrum (photon index of $\Gamma = 4.3$). Over the following months, the X-ray flux started decaying and demonstrated significant flaring activity on both short (hour-) and long (year-) timescales. By inspecting the multiwavelength lightcurves of time-domain wide-field facilities, we detected a strong mid-infrared flare, evolving over 2 years, and a weaker optical counterpart, with possible hints of a rise >3 years prior to the X-ray discovery. Follow-up optical spectroscopy revealed transient features, including redshifted Balmer lines (FWHM $\sim 1500$~km~s$^{-1}$), strong Fe~II emission, He~II, Bowen fluorescence lines, and high-ionization coronal lines such as [Fe~X] and [Fe~XIV]. One spectrum displayed a triple-peaked H$\beta$ line, consistent with emission from a face-on elliptical accretion disk. 
   The spectroscopic features and the slow evolution of the event place J012026 within the nuclear-transient classifications of Bowen fluorescence flares (BFFs) and extreme coronal line emitters (ECLEs). BFFs have been associated with rejuvenated accreting SMBH, although the mechanism triggering the onset of the new accretion flow is yet to be understood, while ECLEs have been linked to the disruption and accretion of stars in gas-rich environments. The association of J012026 to both classes, combined with the X-ray, multi-wavelength, and spectroscopic information, supports the idea that the BFF emission could be, at least in some cases, triggered by tidal disruption events (TDEs) perturbing gaseous environments. The observed short- and long-term X-ray variability, uncommon in standard TDEs, adds complexity to these families of nuclear transients. These results highlight the diverse phenomenology of nuclear accretion events and demonstrate the value of systematic X-ray surveys, such as eROSITA and Einstein Probe, for uncovering such transients and characterizing their physical origin.}

\authorrunning{P. Baldini et al}
\title{A new Bowen Fluorescence Flare and Extreme Coronal Line Emitter discovered by SRG/eROSITA}

\author{P. Baldini\inst{1,}\thanks{corresponding author, \href{mailto:baldini@mpe.mpg.de}{baldini@mpe.mpg.de},},  
  A. Rau\inst{1},
  R. Arcodia\inst{2,}\thanks{NASA Einstein fellow},
  T. Ryu\inst{3,4,5},
  Z. Liu\inst{6},
  P. Sánchez-Sáez \inst{7,8},
  I. Grotova \inst{1},
  A. Merloni \inst{1},
  S. Ciroi \inst{9},
  A. J. Goodwin \inst{10},
  M. Gromadzki \inst{11},
  A. Kawka \inst{10},
  M. Masterson \inst{2}, 
  D. Tubín-Arenas \inst{12},
  D. A. H. Buckley \inst{13,14,15}, 
  F. Mille \inst{16},
  G. E. Anderson \inst{10}, 
  S. Bahic \inst{12},
  D. Homan \inst{12},
  M. Krumpe \inst{12},
  J. C. A. Miller-Jones \inst{10},
  \and  K. Nandra\inst{1}
  } 

\institute{Max-Planck-Institut f\"ur extraterrestrische Physik, Giessenbachstra{\ss}e 1, D-85748 Garching bei M\"unchen, Germany
\and MIT Kavli Institute for Astrophysics and Space Research, 70 Vassar Street, Cambridge, MA 02139, USA 
\and Max-Planck-Institut für Astrophysik, Karl-Schwarzschild-Straße 1, 85748 Garching bei München, Germany 
\and JILA, University of Colorado and National Institute of Standards and Technology, 440 UCB, Boulder, 80308 CO, USA 
\and Department of Astrophysical and Planetary Sciences, 391 UCB, Boulder, 80309 CO, USA 
\and Centre for Astrophysics Research, University of Hertfordshire, College Lane, Hatfield AL10 9AB, UK
\and European Southern Observatory, Karl-Schwarzschild-Strasse 2, 85748 Garching bei München, Germany
\and Millennium Institute of Astrophysics (MAS), Nuncio Monseñor Sotero Sanz 100, Of. 104, Providencia, Santiago, Chile
\and Dipartimento di fisica e Astronomia, Università degli Studi di Padova, Vicolo dell’Osservatorio 3, 35122 Padova, Italy
\and International Center for Radio Astronomy Research, Curtin University, GPO Box U1987, Perth, WA 6845, Australia
\and Astronomical Observatory, University of Warsaw, Al. Ujazdowskie 4, 00-478 Warszawa, Poland
\and Leibniz-Institut für Astrophysik Potsdam, An der Sternwarte 16, 14482 Potsdam, Germany
\and South African Astronomical Observatory, PO Box 9, Observatory Rd, 7935 Observatory, Cape Town, South Africa
\and Department of Astronomy, University of Cape Town, Private Bag X3, Rondebosch 7701, South Africa
\and Department of Physics, University of the Free State, PO Box 339, Bloemfontein 9300, South Africa
\and Las Campanas Observatory – Carnegie Institution for Science, Colina el Pino, Casilla 601, La Serena, Chile
} 

\maketitle
    
\section{Introduction}

When a star wanders too close to a supermassive black hole (SMBH), tidal forces can rip it apart. If the disruption occurs outside the event horizon, as matter falls towards the black hole, a flare of optical/UV/X-ray radiation will appear and last on a timescale of months (\citealp{hills1975possible, rees1988tidal, evans1989, Phinney1989}). These so-called tidal disruption events (TDEs) can illuminate dormant nuclear black holes, complementing the population revealed by active galactic nuclei (AGN). As they predominantly happen around low-mass SMBH or even the elusive intermediate-mass BHs, TDEs are crucial for a complete BH demographic analysis (e.g. \citealp{stone2016}). TDEs are also the perfect environment to study the onset, development, and exhaustion of accretion flows and related ejections, such as jets and outflows. Since their overall evolution is contained within a few years, all their phases can be probed accurately.

The first TDE candidates were discovered in the X-rays with ROSAT (\citealp{trumper1982rosat,komossa1999discovery,komossa1999giant,grupe1999rx, greiner2000rx}), as soft X-ray flares decaying over a few months, following the characteristic $t^{-5/3}$ shape, which can be derived from simple theoretical arguments (see e.g. \citealp{Saxton2020, gezari2021tidal}). In the following years, a few more sources were discovered serendipitously with Chandra, XMM-Newton, and
Swift ( \citealp{weisskopf2000chandra, jansen2001xmm, Gehrels2004}, and see \citealp{komossa2015tidal} and references therein). That said, in the past two decades, most events have been selected in the optical/UV, thanks to the advent of wide-field time-domain surveys, such as the Panoramic Survey Telescope and Rapid Response System (Pan-STARRS, \citealp{Chambers2016}), the Palomar Transient Factory and its successor the Intermediate Palomar Transient Factory (iPTF, \citealp{Law2009, Kulkarni2013}), the All-Sky Automated Survey for Supernovae (ASAS-SN \citealp{shappee2014man,kochanek2017all}), the Asteroid Terrestrial-impact Last Alert System (ATLAS, \citealp{tonry2018atlas}), and the Zwicki Transient Facility (ZTF, \citealp{bellm2018zwicky}).

However, the interplay of optical/UV and X-ray radiation is a crucial element in our attempts to understand the physical origin of the radiation in TDEs. In fact, while it is generally understood that the X-ray emission arises due to accretion processes, it is still unclear where the optical/UV radiation comes from. Moreover, many optically selected TDEs do not show X-ray emission up to $\sim$100 days after optical peak (e.g., \citealp{Guolo2024systematic}). Currently, two main families of models have been proposed: the first proposes that the primary X-ray radiation from the accretion disk is reprocessed to longer wavelengths by a thick envelope (e.g. \citealp{Guillochon2014, roth2016x, metzger2016bright, parkinson2022optical}). In contrast, the second model proposes that the optical/UV emission is due to shocks created by collisions between incoming and outgoing streams on highly eccentric orbits near the apocenter of the returning stream (e.g. \citealp{Piran+2015, shiokawa2015general, jiang2016A, Bonnerot2017, ryu2023shocks, steinberg2024}). In the latter scenario, the accretion disk can only form after the dissipation of orbital energy through the self-crossing shocks, naturally explaining the delayed emergence of X-rays, with respect to the optical flare. The cross-correlation of optical and X-ray lightcurves has provided evidence to support this scenario (e.g. \citealp{guo2023evidence}); however, when early X-ray sampling was available, observations revealed a more complex picture, with the notable example of AT 2022dsb, showing fading X-rays prior to the optical peak (\citealp{malyali2024transient}). Because of this reason, assembling an unbiased sample, with events selected both in the X-rays and in the optical/UV bands is critical for drawing a clearer picture of the physical mechanism behind these transients. 

The systematic selection of events in the X-rays has been greatly enhanced by the extended ROentgen Survey with an Imaging Telescope
Array (eROSITA; \citealp{predehl2021erosita}), the soft X-ray instrument
on board the Spektrum–Roentgen–Gamma (SRG; \citealp{sunyaev2021srg}). By scanning the entire sky once every six months, between 4 and 5 times, eROSITA has provided the largest sample of systematically selected TDEs (13 in \citealp{sazonov2021first}, 6 in \citealp{khorunzhev2022search}, and 31 in \citealp{Grotova2025b}), as well as discovering unique and exotic sources, such as AT 2019avd, eRASSt J133157.9-324321, eRASSt J045650.3-203750, eRASSt J234402.9-352640, AT 2022dsb (respectively, \citealp{at2019avd,malyali2023rebrightening,  liu2023deciphering, homan2023discovery, malyali2024transient}).

Now that the gap between selection techniques is being filled, with over a hundred TDEs discovered, the diversity within these events is becoming increasingly apparent. Some events show only either X-rays or optical/UV variability (e.g. \citealp{Grotova2025b}, \citealp{gezari2012ultraviolet}), while others show repetitions or re-brightenings (e.g. \citealp{malyali2023rebrightening, 2024NatAs...8..347G}), quasi-periodic phenomena (Quasi-periodic Oscillations, QPO, e.g. \citealp{reis2012200, pasham2014400}, Quasi-Periodic Ultrafast outflows, QPOut, e.g. \citealp{pasham2014400}, and Quasi-Periodic Eruptions, e.g. \citealp{miniutti2019nine, giustini2020quasi, arcodia2021}), and a general deviation from a $t^{-5/3}$ behavior (e.g. \citealp{Guolo2024systematic}). Additionally, some other classes of nuclear transients have been connected to the presence of TDEs. 
This is the case for extreme coronal line emitters (ECLE, e.g., \citealp{komossa2008ecle, wang2012ecle}), which show strong and transient high ionization coronal lines in optical spectra, such as [Fe X] $\lambda$6376, [Fe XI] $\lambda$7894, [Fe XIV] $\lambda$5304, and [S XII] $\lambda$7612, which correspond to ionization potentials in the hundreds of eV. Although these lines can also be produced in AGN, ECLEs are distinguished by their extreme ratios of coronal line emission to typical AGN lines, such as [O III] $\lambda$5007, and by the lack of evidence for strong ongoing nuclear activity (e.g., \citealp{cerqueira2021coronal, negus2021physics, prieto2022novel}).
ECLEs are also characterized by optical lightcurves monotonically decaying, similarly to TDEs, but on longer timescales (years vs. months, e.g. \citealp{Newsome2024mapping}).
The transient behavior of both the continuum and the CLs, and rate estimates that align with those of TDEs (\citealp{wang2012ecle}), strongly suggest that ECLEs are powered by TDEs in gas-dense environments, where the flaring UV and soft X-ray radiation can ionize nearby gas, producing the observed line emission (\citealp{Hinkle2024coronal}). Indeed, the emergence of transient coronal lines after the dust-reprocessing echo in the TDE AT 2017gge (\citealp{onori2022nuclear}) first observationally confirmed this scenario, and a similar behavior has been observed in AT 2019qiz (\citealp{short2023}), strengthening the connection.  

Another transient class potentially associated with TDEs is the one identified in \citealp{Trakhtenbrot2019bff} and dubbed Bowen Fluorescence Flare (BFF) in \citealp{makrygianni2023bff}. These events show strong He\,II and N\,III emission, also commonly seen in TDEs (e.g. \citealp{Onori2019, leloudas2019spectral, Blagorodnova2019, vanvelzen2021seventeen}) and produced with the Bowen Fluorescence mechanism (\citealp{1928ApJ....67....1B}), but with widths of the order of $10^3$ km s$^{-1}$, compared to the typical $10^4$ km s$^{-1}$ in TDEs (e.g., \citealp{Arcavi2014}). Similarly to ECLEs, they also decay on overall longer timescales, with flatter power-law slopes, compared to the typical $t^{-5/3}$ (\citealp{Trakhtenbrot2019bff}). BFFs have been associated with 'rejuvenated SMBHs' (\citealp{Trakhtenbrot2019bff}), in which the emission originates from a suddenly illuminated pre-existing broad line region (BLR), where the sudden increase in the extreme UV/X-ray continuum triggers the Bowen Fluorescence mechanism. Indeed, some of these events show clear indication of previous AGN activity (e.g., \citealp{makrygianni2023bff, frederick2021family, Sniegowska2025}), but the rejuvenation mechanism is not clear, and for events AT 2019aalc and AT 2022fpx a TDE connection has been proposed (\citealp{veres2024at2019aalc, koljonen2024at2022fpx}).

In this work, we present the new eROSITA-selected transient eRASSt J012026.5-292727 (hereafter J012026), which was missed by optical surveys and adds a new piece to the puzzle of TDEs, their non-monotonic X-ray lightcurves, gas-rich environments, and the connections with ECLEs and BFFs. The paper is structured as follows: In Section \ref{sec:discovery}, we describe the discovery of J012026, and in Section \ref{sec:x-ray}, we describe the X-ray data, their reduction, analysis, and results. In section \ref{sec:phot} we describe the photometric properties, and in section \ref{sec:optspec} the optical spectroscopy. We present the discussion on the nature of the source in section \ref{sec:discu}, and summarize our results and conclusions in section \ref{sec:sum}. Throughout this paper we adopt a flat $\Lambda$CDM cosmology, with $H_0$ = 67.7 km s$^{-1}$ Mpc$^{-1}$,
$\Omega_m$ = 0.309 (\citealp{planck2016}).

\begin{figure}
    \centering
    \includegraphics[width=0.8\columnwidth]{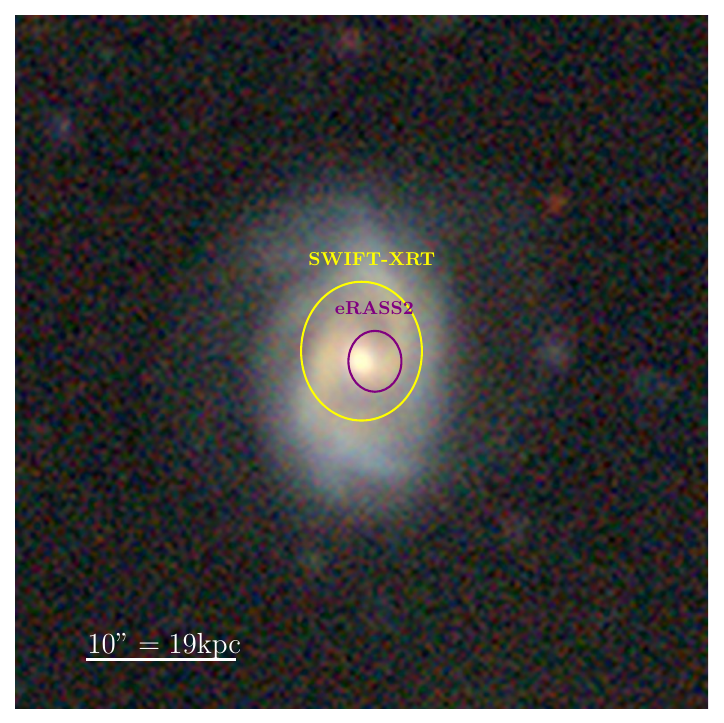}
    \caption{LS10 grz-color-image of the host galaxy of J012026. The purple circle indicates the eRASS2 positional error, and the yellow circle represents the SWIFT-XRT positional error.}
    \label{fig:ls10cutout}
\end{figure}

\begin{figure*}[t]
    \includegraphics[width=\textwidth]{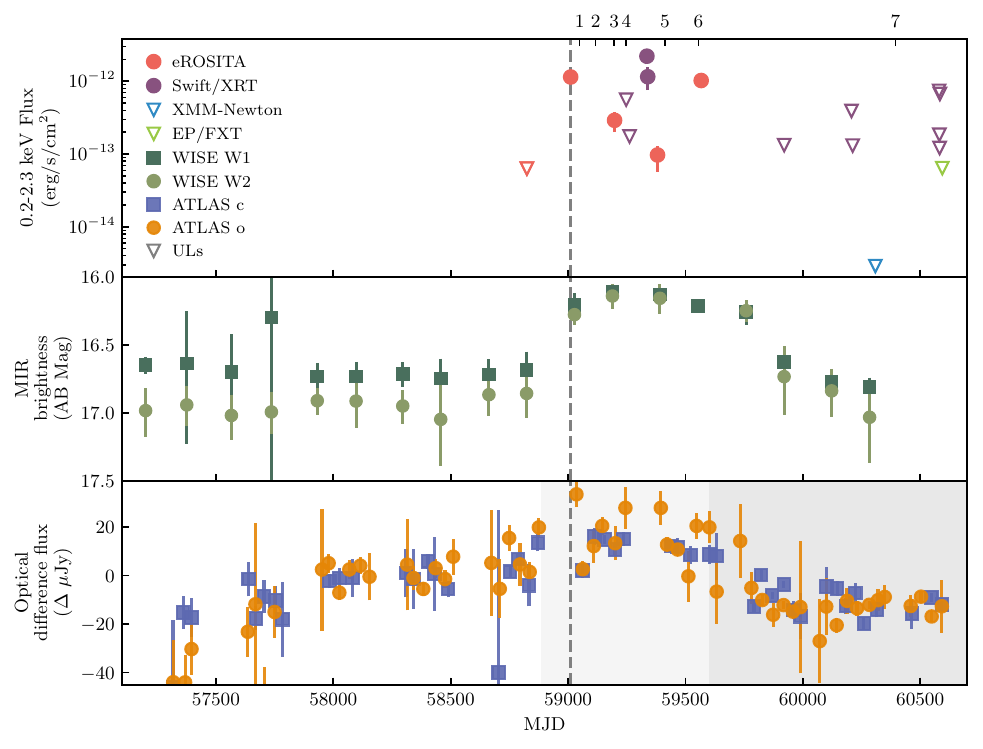}
    \caption{Multiwavelength lightcurve of J012026. The top panel shows the X-ray evolution in the 0.2-2.3 keV band (empty triangles correspond to upper limits). In the middle panel, the WISE W1 and W2 lightcurve is presented in AB magnitudes, and the bottom panel shows the difference-photometry lightcurve for the ATLAS o and c filters in $\mu$Jy. The vertical dashed lines indicate the eRASS2 discovery. The numbered ticks at the top of the plot indicate the epochs at which optical spectroscopy was collected (see \ref{sec:optspec}).}
    \label{fig:pancurve}
\end{figure*}

\section{Discovery and follow-up}
\label{sec:discovery}

\begin{table}[]
\caption{X-ray observation log of J012026. The 2024 Swift/XRT observations are grouped by day. Swift/XRT observations 00014018003 and 00014018004 are referred to in the text as Swift/XRT3 and Swift/XRT4}
\begin{tabular}{llll}
\toprule
Date       & Telescope   & ObsID                                                                     & Exp. {[}s{]} \\ \midrule
2019-12-12           & SRG/eROSITA & eRASS1                                                                    &     245        \\
2020-06-13           & SRG/eROSITA & eRASS2                                                                    &     228         \\ 2020-12-15
           & SRG/eROSITA & eRASS3                                                                    &    253          \\
2021-02-02 & Swift/XRT   & 00014018001                                                               & 2168         \\
2021-02-16 & Swift/XRT   & {\color[HTML]{000000} 00014018002}                                        & 1859         \\
2021-05-01 & Swift/XRT   & {\color[HTML]{000000} 00014018003 (3)}                                        & 1373         \\
2021-05-05 & Swift/XRT   & 00014018004 (4)                                                              & 1063         \\ 2021-06-16
           & SRG/eROSITA & eRASS4                                                                    &    241          \\
 2021-12-19          & SRG/eROSITA & eRASS5                                                                    &  251            \\
2022-12-07 & Swift/XRT   & 00014018005                                                               & 2575         \\
2023-09-20 & Swift/XRT   & 00014018006                                                               & 924          \\
2023-09-24 & Swift/XRT   & 00014018007                                                               & 1978         \\
2023-12-30 & XMM-Newton  & 0923600101                                                                & 53900        \\
2024-09-27 & Swift/XRT   & \begin{tabular}[c]{@{}l@{}}00097365001, ...,  \\ 00097365015\end{tabular} & 4207         \\
2024-09-28 & Swift/XRT   & \begin{tabular}[c]{@{}l@{}}00097365017, ..., \\ 00097365031\end{tabular}  & 5224         \\
2024-09-29 & Swift/XRT   & \begin{tabular}[c]{@{}l@{}}00097365032, ..., \\ 00097365046\end{tabular}  & 4523         \\
2024-10-11 & EP/FXT      & 11908706671                                                               & 1101         \\ \midrule
\end{tabular}

\label{tab:obslog}
\end{table}

J012026 was discovered as a new bright source on 2020-06-13 during the second eROSITA all-sky survey (eRASS2)  (see \citealp{Grotova2025a} for details on the search for non-AGN nuclear transients in eROSITA). The new soft and point-like source was localized at RA=20.1107\deg, Dec=-29.4631\deg, consistent with the nucleus of the z=0.1029 galaxy 2dFGRS TGS295Z163 (\citealp{colless20012df}) (see Fig.~\ref{fig:ls10cutout}). Independently, J012026 was picked up by the blind QPE search algorithm described in \citealp{Arcodia2024cosmic} due to its short-term variability in eRASS5 (see Section \ref{sec:x-ray}).
Following the discovery, several Swift-XRT observations were requested (see Table \ref{tab:obslog}) and the source was sporadically monitored until 2024. To further investigate the short-term X-ray variability, a 53\,ks XMM-Newton observation was later performed, and the source was also monitored every $\sim1$ hour for about 3 days with Swift-XRT in September 2024. Shortly after, an Einstein Probe (EP, \citealp{ep2022}) Follow-up X-ray Telescope (FXT) observation of J012026 was performed as part of an eROSITA TDE follow-up program. The multiwavelength lightcurve of J012026 can be seen in Fig. \ref{fig:pancurve}. 

J012026 was also monitored in a multiwavelength fashion. Several follow-up optical spectra were obtained between 25 days after the discovery and 2024, as discussed in section \ref{sec:optspec}. Moreover, J012026 was observed in radio bands with the Australia Telescope Compact Array (ATCA) in three different epochs, however the detections were marginal (see appendix \ref{app:radio}).

\section{X-rays data and analysis}
\label{sec:x-ray}
\subsection{eROSITA}

J012026 was detected in eRASS2,3,4, and 5, but not in eRASS1. We retreive the eRASS1 upper limits from the eROSITA DR1 upper limit server\footnote{\url{https://erosita.mpe.mpg.de/dr1/AllSkySurveyData_dr1/UpperLimitServer_dr1/}
} (\citealp{tubin2024erosita, merloni2024}). For the remaining all-sky surveys, we use the lightcurves and spectra systematically extracted from the eROSITA processed event files (version 020) with the task \textsc{srctool} of the eROSITA science analysis software (\texttt{eSASS} v.211214.05, \citealp{brunner2022}). We combine light curves and spectra collected from telescope modules (TM) 1, 2, 3, 4, and 6, as TM5 and TM7 are affected by light leaks (\citealp{predehl2021erosita}). The circular extraction region sizes scale with the maxim likelihood (ML) source count rate reported in the catalogs and have radii of 85", 58", 51", and 84", respectively, for eRASS2,3,4 and 5. The background extraction regions are determined as described in \citealp{liu2022efed}.

\begin{figure*}
    \centering
    \includegraphics[width=\textwidth]{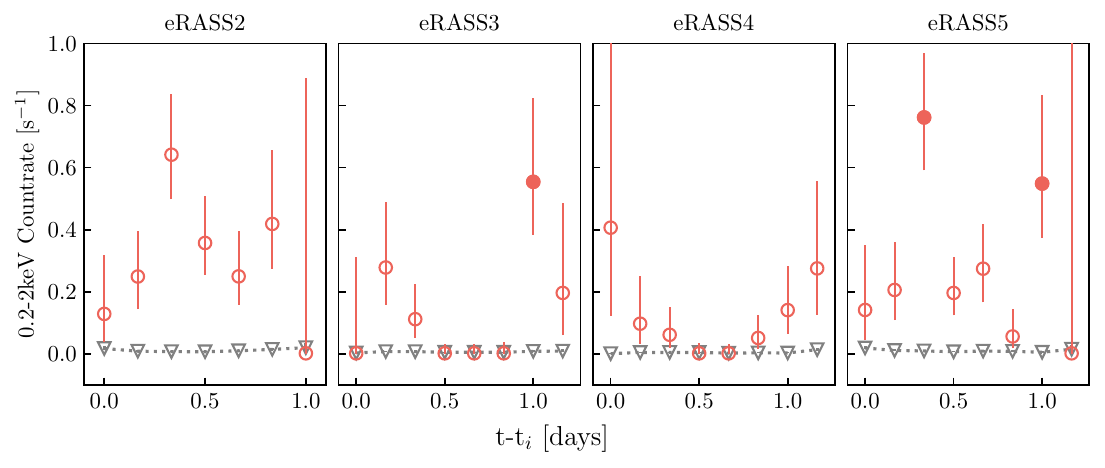}
    \caption{eROSITA eROday 0.2-2.0 keV band lightcurve of J012026. The red, empty circles show the source vignetting and exposure corrected countrates, while the grey triangles correspond to the background countrate. The filled red circles correspond to the eRO-days selected as flaring. The time (x) axis is in units of days since the first observation of the source, for each scan.}
    \label{fig:erolc}
\end{figure*}

The eROSITA scanning strategy is such that each point of the sky is observed repeatedly for up to 40 seconds every four hours (i.e., every eROday, \citealp{predehl2021erosita}). The number of exposures depends on the ecliptic latitude but has a typical value of 6. This implies that for each eRASS, a lightcurve of $\sim1$ day of baseline can be extracted.
The eROSITA eROday X-ray light curve of J012026 is presented in Fig. \ref{fig:erolc}, with the exclusion of eRASS1, as the source is undetected. It can be seen that eRASS3 and eRASS5 show large amplitude eROday variability, with flaring activity present respectively on the 7th (eRASS3), and 3rd and 7th (eRASS5) eROdays. The identification of the flares was initially visual and later confirmed by calculating the significance of the variability through the maximum amplitude deviation (MAD) method (\citealp{boller2016second}). We conservatively compute the amplitude of the variation between the count-rate $C$ of the flaring eROday $(t_{\text{flare}})$ and the eROday with the lowest count rate ($t_{\text{min}}$) in the same eRASS as
\begin{equation}
    \text{MAD} =  \left(C(t_{\text{flare}}) - \sigma_{\text{low}}(t_{\text{flare}})\right) - \left(C(t_{\text{min}}) + \sigma_{\text{upp}}(t_{\text{min}})\right),
\end{equation}

\noindent where $\sigma_{\text{low}}$ and $\sigma_{\text{upp}}$ are respectively the lower and upper $1\sigma$ errors. 
The significance of the variability can be computed as
\begin{equation}
    \text{SIG\_MAD} = 
    \frac{\text{MAD}}{\sqrt{\sigma_{\text{low}(t_{\text{flare}})}^2 + \sigma_{\text{upp}(t_{\text{min}})}^2}}.
\end{equation}

As shown in \citealp{buchner2022systematic}, the value of $\text{SIG\_MAD}$ is an underestimation of the true significance of the variability. In their work, they report values of $\text{SIG\_MAD}$ corresponding to 3$\sigma$ significance at a given average count rate. Based on Fig. 13 of \citealp{buchner2022systematic}, we adopt a $3\sigma$ threshold of 1.5 for J012026. We measure $\text{SIG\_MAD}=1.6$ for eRASS3, and 2.4 and 2.0, respectively, for the first and second eRASS5 flares. 

We analyzed eRASS2-5 spectra with the Bayesian X-ray Analysis software (\texttt{BXA}) version 4.1.2 (\citealp{buchner2014bxa}), which connects the nested sampling algorithm UltraNest (\citealp{buchner2019ultranest, buchner2021ultranest}) with the fitting environment CIAO/Sherpa (\citealp{fruscione2006sherpa}). Spectra were fit unbinned and using C-statistic. The fitting procedure included a PCA-based background model (e.g., \citealp{simmonds2018}) derived from a large sample of eROSITA background spectra (\citealp{liu2022efed}). 
For all spectra, we fit both a multi-color disk model (\textsc{diskbb}) and a power-law model (\textsc{powerlaw}), as shown in Fig. \ref{fig:xrayspeca}. In both cases, we included the contribution of Galactic absorption through the model \textsc{tbabs}, which is due to a line-of-sight column density of $N_{H}=1.44\times10^{20}$ cm$^{-2}$ , as estimated by the HI4PI collaboration (\citealp{Hi4pi}).


In order to test a potential spectral evolution in conjunction with the flares in eRASS3 and eRASS5, we repeated the spectral analysis by first separating the flares from the rest of the respective eRASS. We isolated the 3rd eROday in eRASS3, and the 3rd and 7th eROday in eRASS5 with the eSASS task \textsc{evtool}. We then combined the 2 eRASS5 flares into a single spectrum, to maximize counts, and then repeated the fitting procedure four times, respectively on the eRASS3 flare, the eRASS5 flares, and the remaining "baseline" counts in eRASS3 and in eRASS5.   

The results of our fit are reported in Table \ref{tab:ero_spec}, and the spectra can be visualized in Appendix \ref{xrayspec}.
Both models reproduce the data well for all observations, although a comparison of Bayesian evidence suggests that a power-law model is preferred in eRASS2 and for the eRASS3 flare. However, the evidence is not decisive (see \cite{kass1995bayes}), and a visual inspection of the spectra in appendix \ref{xrayspec} does not show a significant difference between the two models. 
eRASS4 shows a drastically different spectral shape compared to the other observations. Since this observation corresponds to the lowest flux state, with a total of 12 source+background counts in the 0.2-2.3 keV band, we do not consider this variation to be physical. 

Excluding eRASS4, there is no drastic spectral variation among eRASSes. However, there is mild evidence for a general harder-when-brighter behavior, which will be discussed jointly with the Swift-XRT observations in the next section. 

We note that J012026 is located at a distance of 60$"$ from the known AGN WISEA J012025.95-292637.3. The latter was detected by the eROSITA pipeline only in eRASS1 \citealp{merloni2024}), while it was not found in later epochs (See also Fig. \ref{fig:cutouts}). In Appendix \ref{sec:qso}, we show through a more conservative extraction procedure that, if present, contamination from WISEA J012025.95-292637.3 is negligible.

\subsection{Swift - XRT}
\begin{table*}[t]
 \caption{Spectral fit results for the eROSITA and Swift-XRT observations of J012026. In the second column we indicate whether the results correspond to the full spectrum, the flaring eROdays only, or the remaining baseline eROdays. Based on the difference in Bayesian evidence (last column), we indicate which of the two models is preferred in the third column. We report both the \textsc{diskbb} temperature kT and the photon index $\Gamma$ for all spectra in the 4th and 5th columns, as well as the 0.2-2.3 keV flux derived for each model respectively in the 6th and 7th column.}
    \footnotesize
    \setlength{\tabcolsep}{3.5pt}
    \centering
    \resizebox{\textwidth}{!}{
        \begin{tabular}{cccccccccc}
            \toprule
            \multicolumn{1}{c}{Epoch} &
            \multicolumn{1}{c}{Spectrum} &
            \multicolumn{1}{c}{Best model} &
            \multicolumn{1}{c}{kT [eV]} &
            \multicolumn{1}{c}{$\Gamma$} &
            \multicolumn{1}{c}{$F^{\rm disk}_{\rm 0.2-2.3\,keV}$ [erg\,s$^{-1}$\,cm$^{-2}$]} &
            \multicolumn{1}{c}{$F^{\rm p-law}_{\rm 0.2-2.3\,keV}$ [erg\,s$^{-1}$\,cm$^{-2}$]} &
            \multicolumn{1}{c}{$\Delta \log Z$} \\
            \midrule
            eRASS2 & Full & \texttt{power-law} & $90\pm10$ & $4.35^{+0.34}_{-0.30}$ & $9.3^{+1.6}_{-1.4} \times 10^{-13}$ & $1.14^{+0.15}_{-0.18} \times 10^{-12}$ &  $4$ \\
            \midrule
            eRASS3 & Full & \texttt{--} & $80^{+20}_{-10}$ & $4.79\pm0.64$ & $2.4^{+0.9}_{-0.7} \times 10^{-13}$ & $2.9^{+1.0}_{-0.9} \times 10^{-13}$ & 0 \\
            & Baseline & \texttt{--} & $80\pm10$ & $5.20^{+0.5}_{-0.57}$ & $4.3^{+2.5}_{-1.6} \times 10^{-14}$ & $4.5^{+2.5}_{-1.3} \times 10^{-14}$ & 0 \\
            & Flare & \texttt{power-law} & $90\pm10$ & $4.29^{+0.96}_{-0.95}$ & $1.17^{+0.61}_{-0.47} \times 10^{-12}$ & $1.70^{+1.02}_{-0.72} \times 10^{-12}$ & 2 \\
            \midrule
             Swift-XRT3 & Full & \texttt{power-law} & $170^{+20}_{-12}$ & $3.66\pm0.31$ & $1.4^{+0.4}_{-0.3} \times 10^{-12}$ & $2.3\pm0.5 \times 10^{-12}$ & $0$ \\
            \midrule
             Swift-XRT4 & Full & \texttt{power-law} & $190^{+39}_{-32}$ & $3.53^{+0.47}_{-0.46}$ & $6.9^{+2.1}_{-1.6} \times 10^{-13}$ &  $1.4^{+0.2}_{-0.3} \times 10^{-12}$ &  $4$ \\
            \midrule
            eRASS4 & Full & \texttt{--} & -- & $2.07^{+0.64}_{-0.68}$ & -- & $9.7^{+0.3}_{-0.4} \times 10^{-14}$ & -- \\
            \midrule
            eRASS5 & Full & \texttt{--} & $100\pm10$ & $4.21^{+0.31}_{-0.27}$ & $(8.1\pm1.1) \times 10^{-13}$ & $1.02^{+0.18}_{-0.14} \times 10^{-12}$ & 0 \\
            & Baseline & \texttt{--} & $70\pm10$ & $5.28^{+0.48}_{-0.57}$ & $5.8^{+1.9}_{-1.6} \times 10^{-13}$ & $7.8^{+2.7}_{-2.4} \times 10^{-13}$ & 0 \\
            & Flare & \texttt{--} & $120\pm20 $ & $3.97\pm0.48$ & $1.45^{+0.43}_{-0.33} \times 10^{-12}$ & $2.08^{+0.55}_{-0.65} \times 10^{-12}$ & 0 \\
            \midrule
        \end{tabular}
    }
   
    \label{tab:ero_spec}
\end{table*}

J012026 was observed with Swift/XRT seven times before 2024 and every hour for three days in 2024 (see Table \ref{tab:obslog}), all in photon-counting mode. The XRT light curves and spectra were generated and downloaded from the UK Swift Science Data Centre website (\citealp{evans2007online,evans2009methods}). We bin the lightcurves before 2024 by snapshot, while for the observations taken in 2024, we bin by 1 day to derive meaningful upper limits. The source was detected only in observations 00014018003 and 00014018004 (Swift-XRT3 and Swift-XRT4 hereafter; see Table \ref{tab:obslog}). We compute flux upper limits for all other observations by assuming a diskbb model with T$_{in}$=100 eV, assuming that the emission is due to an accretion disk (see discussion in Section \ref{sec:discu}).

We analyze the spectra of Swift-XRT3 and Swift-XRT4 following the same prescription as for the eROSITA data. We report the results in Table \ref{tab:ero_spec}, and we note that the peak luminosity of the Swift/XRT detections is similar to the eROSITA flares. In Swift-XRT4 the power-law model is statistically more favored to the diskbb, but, similarly to the eROSITA observations, a visual inspection does not show any particular preference (See Fig \ref{fig:xrayspeca}). 

\begin{figure}
    \centering
    \includegraphics[width=\columnwidth]{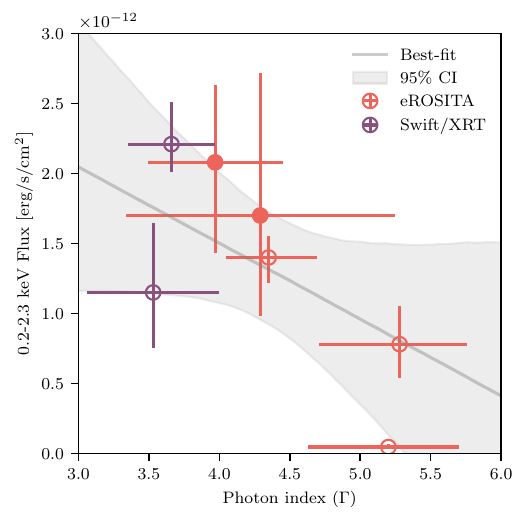}
    \caption{0.2-2.3 keV flux vs spectral photon index for X-ray detections. Filled red markers correspond to eROSITA flaring states, while empty markers to the baseline spectra. We do not include eRASS4, due to its low flux level.}
    \label{fig:hwb}
\end{figure}

As mentioned in the previous section, the eROSITA and Swift-XRT spectral analysis reveal a general harder-when-brighter behavior. In Fig. \ref{fig:hwb}, we plot the 0.2–2.3keV flux versus the photon index $\Gamma$ for eRASS2, the eRASS3 and eRASS5 baselines and flares, and Swift-XRT3 and 4. We overplot a linear fit, which suggests an anticorrelation; however, the 95\% confidence interval, derived from Monte Carlo iterations accounting for uncertainties in both variables, is consistent with no correlation, as shown by the grey band. We also compute the distribution of the Pearson correlation coefficient in an analogous Monte Carlo fashion, and confirm the lack of statistical significance of the anticorrelation (The details of this test are presented in Appendix \ref{xrayspec}). 

\subsection{XMM-Newton}

XMM-Newton observed the location of J012026 as a target of opportunity (ToO) on December 31st, 2023 (See Table \ref{tab:obslog}). No source was detected at the position of J012026.
The Observation Data Files (ODFs) were reduced using the XMM-Newton Science Analysis System software (\texttt{SAS}, version 21.0, \citealp{gabriel2004xmm}), with the latest available calibration files. We generated the event lists for the
European Photon Imaging Camera (EPIC) MOS (\citealp{turner2001european}) and pn (\citealp{struder2001european}) detectors through the SAS tasks \textsc{emchain} and \textsc{epchain}. We identified periods of high background flaring and filtered the event lists accordingly, resulting in an effective exposure time of 40215s. 
We extracted counts at the J012026 2dF source position for all three cameras with circular regions with radii of 20" and background counts from circular regions with radii of 45". We estimate the 3-$\sigma$ flux upper limit by calculating the Poissonian confidence interval and assuming a \textsc{diskbb} model with $T_{in}$=100\,eV, consistently with the Swift-XRT upper limits. This result in a 3-$\sigma$ upper limit in the 0.2-2.3 keV band of $f<3.3 \times 10^{-15} \mathrm{ergs/s/cm^2}$

\subsection{Einstein Probe - FXT}

The 2024-10-11 EP/FXT observation yielded no detection (see Table \ref{tab:obslog}). We used the pipeline \textsc{fxtchain} of the FXT Data Analysis Software (\texttt{FXTDAS}) v.1.1, to produce clean event files. We derived upper limits for the two telescope modules FXT-A and FXT-B (\citealp{yuan2025science}) by extracting photon counts from a source circular region with radius 30", a background region ten times larger, and assuming the same \textsc{diskbb} model with kT=100\,eV. The respective FXT-A and FXT-B  3$-\sigma$ flux upper limit in the 0.2-2.3 keV band are $f<2.3 \times 10^{-14} \mathrm{ergs/s/cm^2}$ and $f<3 \times 10^{-14} \mathrm{ergs/s/cm^2}$.

\section{Photometry}
\label{sec:phot}
\subsection{Mid-infrared variability}

The location of J012026 has been observed since 2013 twice per year as part of the NEOWISE reactivation mission (NEOWISE-R; \citealp{mainzer2014}). We obtained the NEOWISE-R light curve from the NASA/IPAC Infrared Science Archive (IRSA), using the Table Access Protocol (TAP) service\footnote{\url{https://irsa.ipac.caltech.edu/TAP/}} by compiling all source detections within 2$"$ of the position of the host-galaxy nucleus. We rebinned individual flux measurements to one point every six months for all-sky scans and converted them into magnitudes (see Fig. \ref{fig:pancurve}, middle panel). 
The historic MIR lightcurve of J012026 was flat until 58823 but exhibited a significant brightening of  $\sim$0.5\,mag in $W1$ and $\sim$0.7\, mag in $W2$ at the time of the eRASS2 detection (marked with a vertical dashed line in Fig.\ref{fig:pancurve}). The lightcurve plateaued for $\sim$2 years until MJD 59758, after which it decayed back to the quiescent level. The pre-flare $W1-W2$ color was -0.26 mag (0.4 in the Vega system) but reddened to -0.04 mags (0.6 Vega mags) during the outburst, closer to the AGN color selection (e.g. 0.8 in the Vega system, \citealp{stern2012mid, assef2013mid})

\subsection{Optical variability}

We searched for an optically variable counterpart by obtaining the Asteroid Terrestrial Impact Last Alert System (ATLAS) lightcurve through the forced photometry service\footnote{\url{https://fallingstar-data.com/forcedphot/}} (\citealp{tonry2018atlas, smith2020design, shingles2021release}). ATLAS uses two 0.5-m telescopes in Hawaii (Haleakala and Mauna Loa Observatory) to cover roughly a quarter of the sky per night, obtaining images in two broadband filters, \textit{c} and \textit{o}, covering respectively 420–650 and 560–820 nm. We present the difference PSF photometry in Fig. \ref{fig:pancurve}, bottom panel, binned by 45 days to obtain meaningful detections, by adapting the software provided by \citealp{Young_plot_atlas_fp}\footnote{\url{https://github.com/thespacedoctor/plot-results-from-atlas-force-photometry-service}}.

We notice that the source significantly increased in flux before MJD 58000. Moreover, close to the eRASS2 detection, we observe a significant rise, especially in the o band. Lastly, the emission decays and then plateaus on timescales comparable to the MIR flare. 
We note that the reference template image used to perform difference photometry is periodically replaced by the forced photometry service. The epochs corresponding to different templates are marked in Fig. \ref{fig:pancurve} with different shades of grey. Since the change of templates occurred during the rise and during the decline of the optical transient, as retrieved from the light-curve files, it is not possible to determine whether the plateau is at a level compatible with pre-58000 or with the 58000-eRASS2 level.

\subsection{Host galaxy properties}

As shown in Fig. \ref{fig:ls10cutout}, the host-galaxy of J012026 is a spatially resolved, face-on spiral galaxy with possible hints of a bar. In order to derive its properties, we compiled the MIR-to-near UV spectral energy distribution (SED) of J012026 using the tool RainbowLasso\footnote{\url{https://github.com/JohannesBuchner/RainbowLasso}}, released in conjunction with the SED-fitting code Genuine Retrieval of AGN Host Stellar Population \texttt{GRAHSP} (\citealp{buchner2024genuine}). The tool retrieves near-UV magnitudes from GALEX (\citealp{bianchi2017revised}), and optical ($g$, $r$, $i$, $z$) and WISE ($W1-W4$) fluxes from Legacy Survey data release 10 (LS10, \citealp{dey2019overview}). Given the large extent of the host galaxy of J012026, we chose to use aperture photometry. After visually inspecting the azimuthal brightness profiles, we select apertures of 7$"$ for the DECam filters $g$, $r$, $i$, and $z$, and 9$"$ for the $W1-W4$ filters. For GALEX, we take the total flux. 

\begin{figure*}
    \centering
    \includegraphics[width=\textwidth]{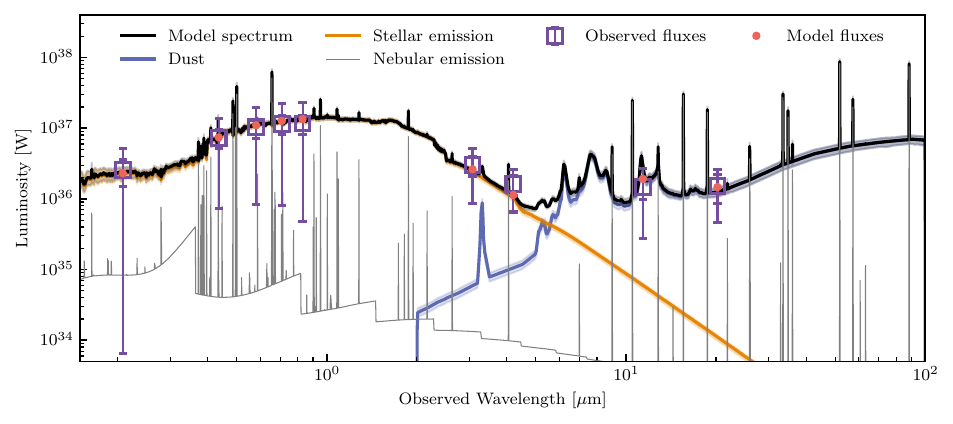}
    \caption{Observed and modeled SED of J012026. The purple squares represent, respectively, the GALEX, DECam $g$, $r$, $i$, $z$ and $W1-W4$ observed fluxes, while the red dots represent the model predicted fluxes. The orange, grey, and blue lines represent the best fit individual model components, which are respectively stellar, nebular, and dust emission. The black line is the total best-fit model.}
    \label{fig:sed}
\end{figure*}

We then model the SED using \texttt{GRAHSP}, including both stellar and nebular components, attenuated by dust and redshifted.  The best-fit solution corresponds to a stellar mass of $log(M_{*}/M_{\odot}) = 11\pm0.1$ and a star formation rate (SFR) of SFR= $5 \pm 1 M_{\odot}\ yr^{-1}$. By using the scaling relations presented in \citealp{reines2015relations}, we obtain a black hole mass of 10$^{7.5\pm0.5}$ M$_{\odot}$. The observed and modeled SED is shown in Fig. \ref{fig:sed}.

\section{Optical spectroscopy}
\label{sec:optspec}

\subsection{Spectroscopic observations}

\begin{table}[]
\caption{Summary of the spectroscopic observations of J012026. For follow-up observations we also report time since the eRASS2 discovery.}
\begin{tabular}{llllc}
\toprule
ID & UT date  & Tel.     & Inst. & Exp. {[}s{]} \\ 
\midrule

0 & 2001-09-14 & AAT      & 2dF        & 1200                  \\ 
\midrule
1 & 2020-07-18 (+25d) & SALT     & RSS        & 1200                  \\
2 & 2020-09-24 (+105d) & ANU 2.3m & WiFeS      & 2700                  \\
3 & 2020-12-12 (+184d) & Baade    & IMACS      & 300                   \\
4 & 2021-02-02 (+236d) & Clay     & LDSS3      & 300                   \\
5 &2021-07-17 (+402d) & SALT     & RSS        & 1800                  \\
6 & 2021-12-06 (+544d) & NTT  & EFOSC      & 1800                  \\
7 & 2023-12-16 (+1283d)  & Baade  & MagE       & 900                   \\ 
\midrule
\end{tabular}

\label{tab:opt_spec_log}
\end{table}

Seven optical spectroscopic observations were performed, starting 25 days after the eRASS2 discovery, and lasting $\sim$ four years. Additionally, an archival spectrum taken as part of the 2dF survey in 2001 is available. A log of the observations is presented in Table \ref{tab:opt_spec_log}, and a description of the data reduction is presented in Appendix \ref{app:opt}.
Both of the SALT/RSS spectra have been observed in long-slit mode, with a resolution of R$\sim1200$ at 5600$\AA$. The WiFes observations are performed using the R3000 and B3000 gratings, which corresponds to a spectral resolution of R$\sim3000$ in both the blue and red arm. For the Baade IMACS observation, we used the Grism 300 with a slit of 0.7$"$, and for the Clay LDSS3 observation, we used the VPH-All grating with a 1$"$ slit. For the NTT/EFOSC2 observations, we used the grating Gr\#13 with a 1.2$"$ slit. Lastly, for the Badee/MagE observations, we used a 0.5$"$ slit. In the following text we refer to the observation by the ID number reported in table \ref{tab:opt_spec_log}. 

\begin{figure*}
    \centering
    \includegraphics[width=\textwidth]{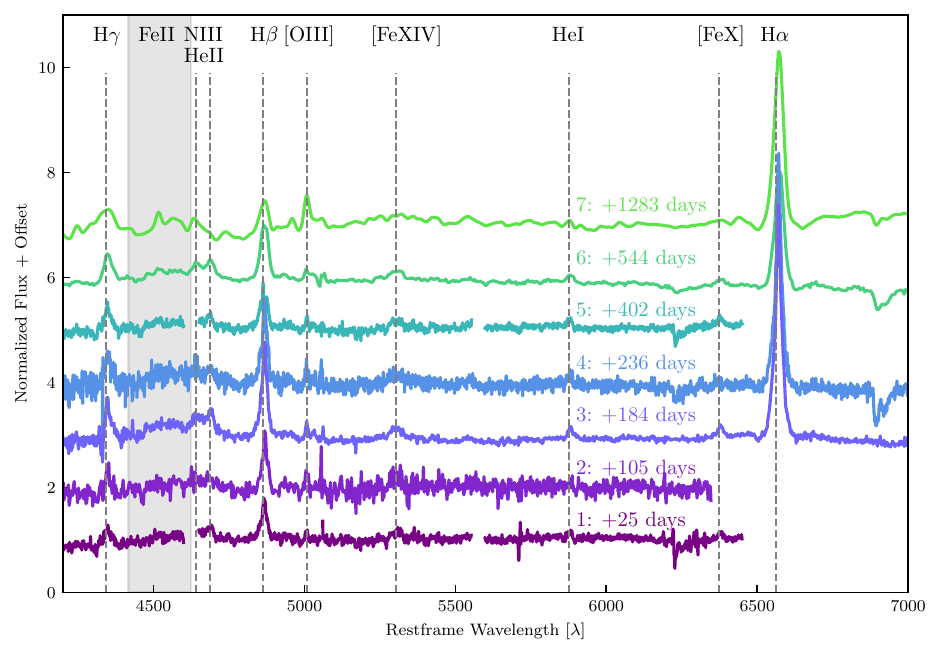}
    \caption{Normalized rest-frame follow-up spectra of J012026, offset by an arbitrary factor for visualization purposes. The most significant emission lines are indicated with dashed lines, and the grey-shaded area shows the prominent Fe II complex. Above each spectrum we report the ID, and phase since eRASS2 discovery in days.}
    \label{fig:allspectra}
\end{figure*}

We recomputed the redshift of J012026 by fitting two Gaussian functions to the Ca\,{\sc ii} H \& K absorption doublet for the two SALT/RSS spectra (1 and 5), and by fitting one Gaussian function to the [OIII] emission line in spectra 1, 3, and 5. We do so by using the \texttt{lmfit} Python package, which employs a Levenberg-Marquardt algorithm to fit the data (\citealp{newville2016lmfit}). The results converge to a redshift of $z$=0.1029. The rest-frame optical spectra of J012026 are shown in Fig. \ref{fig:allspectra}. Interestingly, J012026 is characterized by a number of peculiar spectral features, such as: 1) redshifted Balmer lines with FWHM$\sim$1500 km s$^{-1}$, 2) a strong Fe\,II complex, 3) He\,II and N\,III lines, 4) high ionization coronal lines 5) a triple-peaked \hb. In the following sections, we accurately describe the analysis performed on these observations and the characterization of the features. The implications of our results are later discussed in section \ref{sec:discu}.  

\subsection{Continuum removal}

Given that the spectra were obtained with different instruments and processed following different prescriptions, we first normalize them to their restframe 5900-6200$\AA$ median flux. We chose this wavelength range as it is covered by all spectra, it is line-free, and it is less affected by the flare emission, which is more prominent at bluer wavelengths. We do not apply any Galactic reddening correction, as the color excess measured towards J012026 has a value of E(B-V)=0.015, as obtained from the dust maps of \citealp{Schlafly2011}. This would correspond to a flux correction $<5\%$, which we consider negligible. 

We then remove the continuum contribution in order to model the residual emission lines. The continuum emission is the combination of the host galaxy, a strong Fe II complex emission, and a blue excess due to the transient flare. We subtract the host contribution by using the archival 2dF spectrum, calibrated using the average response function from \citealp{lewis2002anglo}. We then model the spectrum with the code \texttt{pPXF} (\citealp{Cappellari2023}), using the stellar population synthesis models of \citealp{conroy2010fsps}. By doing so, we are able to interpolate in a physically motivated fashion the low-resolution 2dF spectrum to the higher resolution follow-up spectra, and to remove the contributions of the artifacts. The subtraction is performed after interpolating the 2dF template to the target spectrum resolution and minimizing the normalized flux difference in the 5900-6200 range. An example of the procedure is shown in the left panel of Fig. \ref{fig:continuumremoval}.

After subtracting the host, we fit each spectrum with a power-law in the line-free regions 4000-4100 $\AA$, 5400-5500 $\AA$, and 5900-6200 $\AA$. For each spectrum, both normalization and slope are left free to vary. After removing the contribution of the powerlaw, we fit the Fe II template presented in \citealp{park2022new} on the residual spectrum over the whole 4000-5600 range. The Fe II template is convolved with a Gaussian kernel, of which the width is left as a free parameter to allow for line broadening. The normalization and line centroids of the Fe II template are left as free parameters. After a best-fit solution is found and the results are visually inspected, we subtract the Fe II template. An example of this procedure is shown in the right panel of Fig. \ref{fig:continuumremoval}.

\subsection{Emission line analysis}

\begin{figure*}[!t]
    \centering
   \includegraphics[width=\textwidth]{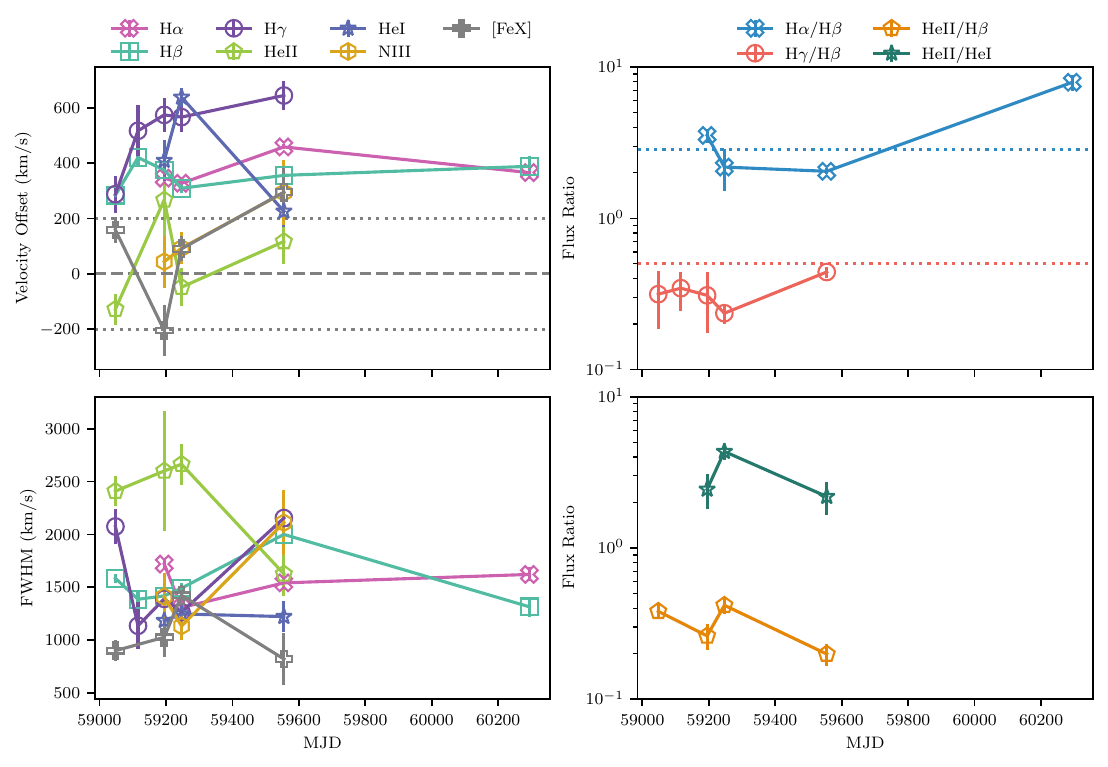}
   \caption{Line parameter evolution of J012026. In the top right panel, we plot the limit on the velocity resolution of the follow-up spectra with dotted lines. In the top right panel, the dotted lines indicate the theoretical values of an unobscured system of the corresponding color ratios.}
   \label{fig:opt_spec_results}
\end{figure*}

We model residual emission lines with a combination of Gaussian functions by using the \texttt{lmfit} Python package, as for the redshift estimation. For each spectrum, we first visually identify which lines are present, and later we fit them with one to four Gaussian, depending on how significantly the chi-square improves when adding one more component. In order to quantify the systemic shift and width of the lines in a consistent way, we use the centroid and FWHM resulting from the single Gaussian fit, while to fully characterize the flux, we combine the contribution of all significant Gaussians. Two examples of this fitting procedure can be seen in Fig. \ref{fig:linefitting}.

Spectrum 1 shows clear \hb and \hg emission (\ha is out of the observed spectral range) along with He\,I 5876 $\AA$, the high-ionization 'coronal' lines [Fe\,X] $\lambda$6375 and [Fe\,XIV] $\lambda$5303, He\,II $\lambda$4686 and [OIII] $\lambda$5007. 
We note here that the Balmer lines are significantly redshifted with respect to the emission from the other lines and, specifically, from the [OIII]. This feature will stay consistent in all spectra and it will be carefully discussed in Section \ref{sec:discu}. Spectrum 2 has a higher resolution but lower S/N ratio. Therefore, we re-bin it by a factor of 4. We observe the same line as spectrum 1 below 5500 $\AA$ (He\,I is not detected, and [Fe\,X] is outside of observed the spectral range), and tentatively detect the presence of the N\,III $\lambda$4641 emission complex. 
We also note that the [Fe\,XIV] appears particularly pronounced and broad. This could be due to a blending with [Fe\,VII] $\lambda$5276. In fact, in all spectra, the FWHM of the [Fe\,XIV] line is $>4000$ km s$^{-1}$, a factor $>2$ compared to all other emission lines, and it is centered at 5292$\pm$2 $\AA$, in between the central wavelengths of [Fe\,VII] and [Fe\,XIV]. However, we note that the S/N of this feature in all spectra is not high enough to allow for an accurate analysis. 
Spectrum 3, extending to longer wavelengths, is also able to reveal the presence of a strong \ha line in addition to the aforementioned features, which are also detected in the following spectrum 4. In addition, in spectra 3 and 4 we confirm the detection of N\,III. In spectrum 5, we observe a unique feature that is not present in any other spectra, namely a triple-peaked \hb, while still noting the presence of all aforementioned features. We discuss the modeling and interpretation of this feature in Section \ref{sec:tphb}. Spectra 6 and 7 appear similar to the other spectra covering the \ha range, although spectrum 7 has a lower S/N, and due to the rebinning, we can no longer confidently identify [Fe\,X], [Fe\,XIV](+[Fe\,VII]), He\,I and He\,II. We also note that the \hb amplitude is now comparable with the [OIII].

In a more quantitative fashion, we report the fit results for all spectra in Table \ref{tab:alllines}, and we plot the results of our analysis in Fig. \ref{fig:opt_spec_results}. In the two left panels, we show the temporal evolution of the line velocity offset (\textit{top panel}) and FWHM (\textit{bottom panel}) of the main modeled lines, excluding spectrum 5 because of the triple-peaked \hb. In the two right panels, instead, we plot the evolution of emission line ratios. Because of the potential blending, we exclude the [Fe\,XIV]+[Fe\,VII] feature from this analysis.

From the top left panel, we can observe that all Balmer lines are redshifted with respect to the [OIII] expected rest-frame wavelength at all epochs. The \hg velocity offset appears to vary between the first and second epoch from 300 km s$^{-1}$ to 500 km s$^{-1}$, but we argue that this could be an effect of Fe\,II subtraction, which affects this emission line the most. In general, the shift of the Balmer lines is between 300 and 600 km s$^{-1}$ and is also seen in the He\,I line. The He\,II, N\,III, and [Fe\,X] instead, are always consistent with a shift of less than 200 km s$^{-1}$ within the errors, a lower limit on the velocity resolution of our spectra, making this consistent with no systemic offset, and suggesting the presence of two separate kinematic components.

The bottom left panel, showing the FWHM evolution for the same lines, supports the kinematic disjunction of the He\,II with respect to the lower ionization energy Balmer, He\,I and N\,III lines. We again argue that the apparent variations of the \hg profile in the first epoch is likely due to the degeneracy of the emission line with the Fe\,II complex. Additionally, we also argue that the anti-correlated variation in the He\,II, and N\,III profiles in spectrum 6 could also arise due to the low wavelength separation of the two features, which adds a degree of degeneracy. In fact, due to the modest resolutions of all of the spectra, we generally warn to interpret with caution the relative parameters in the blended  He\,II, and N\,III complex. 

The top right plot shows the evolution of the \ha/\hb 
 and \hb/\hg Balmer ratios. With the dotted horizontal lines, we show the intrinsic value of the corresponding color ratio for a photoionized gas with density $n_e=10^2-10^4$ cm$^{-3}$ and temperature $T=10^4 \ K$, as set by atomic physics assuming case B recombination (\citealp{1989agna.book.....O}). For all spectra but the last, the observed ratios are generally consistent with theoretical predictions or possibly lower, indicating that photoionization might not be the only responsible process for the production of the emission lines or that the density and temperature assumptions might be too simplistic. However, the increase of the \ha/\hb ratio to 7.9 in the latest spectrum, which is 2.7 times larger than the expected value of 2.86, is a clear indication of late-time reddening.  

 Finally, in the bottom right plot, we show ratios of He\,II over \hb or He\,I in order to check consistency with TDE models, which invoke a receding photosphere and predict these ratios to increase over time (\citealp{roth2016x, charalam2023}). As can be noted by the figure, we do not observe any such trend.

\subsection{Modeling of the triple-peaked H$_\beta$}
\label{sec:tphb}

As introduced in the previous section, spectrum 5, taken 402\,days after eRASS2 discovery with SALT/RSS, revealed a \hb profile significantly different from all other observations. This is because the line appears triple-peaked, with the central peak coincident with the expected rest-frame wavelength at z=0.1029, differently from all other spectra, in which the line is peaking significantly redward (see Fig. \ref{fig:linefitting}, bottom). This type of profile has been observed before in TDEs (AT 2018hyz \citealp{short2020}, PS18kh \citealp{holoien2018,hung2019}, AT 2020zso \citealp{wevers2022}, AT 2020nov \citealp{earl20242020nov}) and in the spectra of AGN (e.g., \citealp{ochmann2024, ward2024panic}), and theoretical models have explained these profiles as the combination of a central Gaussian emission line, and a double-peaked, broad line produced within an elliptical accretion disk (\citealp{eracleous1995}, E95). Motivated by this, we explore which parameter space can reproduce the spectrum by implementing the E95 model.

\begin{figure}[t]
    \centering
    \includegraphics{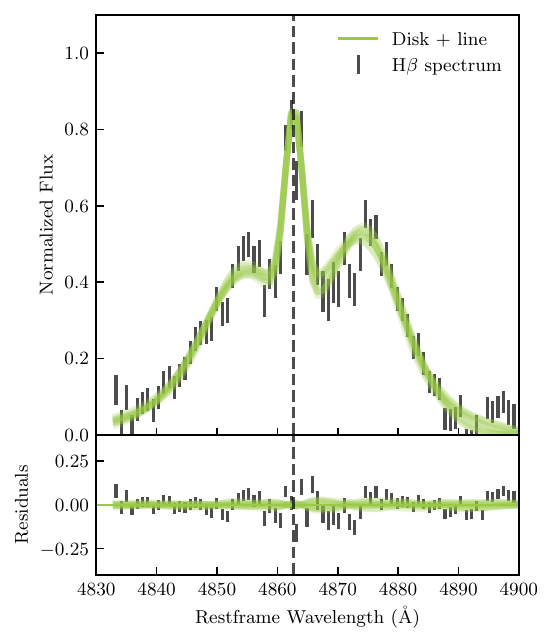}
    \caption{Modeling of the triple-peaked \hb line with a narrow gaussian plus the E95 elliptical disk model. The green band shows the best fit model, and the vertical dashed line corresponds to the restframe centroid of \hb.}
    \label{fig:triplehbeta}
\end{figure}

The E95 model has seven parameters: the inner and outer pericenter disk radii $\xi_1$ and $\xi_2$ in units of gravitational radii $R_g= GM_{SMBH}/c^2 $, the emissivity profile $q$, which scales as $\xi^{-q}$, with $\xi_1 <\xi < \xi_2$, the inclination of the disk plane with respect to the line of sight $i$, the intrinsic velocity broadening $\sigma_{disk}$, the disk eccentricity $e$ and the disk azimuthal angle with respect to the apocenter $\phi_0$. 
We implemented the model following the formalism presented in E95, where the observed line profile is computed by integrating the emitted flux from elliptical disk elements. We create a $7^7$ grid of synthetic spectra resulting from the combination of parameters in the following range: $100 \ R_{g} < \xi_1 < 3000 \ R_{g}$, $10000 \ R_{g} < \xi_2 < 100000 \ R_{g}$, $ 0.5 < q < 2 $, $100 \ \mathrm{km \ s^{-1}} <\sigma_{disk}< 500 \ \mathrm{km s^{-1}}$, $10 \deg< i < 80 \deg $, $0.5 < e < 0.9$, and $0 \deg< i < 360 \deg $.
We fit the model by minimizing the $\chi^2$ between the observed H$\beta$ profile and a composite model consisting of the synthetic E95 profile and a narrow Gaussian model (FWHM $<250$ km s$^{-1}$), with centroid free to vary between the restframe wavelengths 4859 and 4863 Å, to account for the central peak. The modeling and fitting were performed through our custom python routine, which extensively made use of the \texttt{SciPy} module (\citealp{scipy}). The best-fit model, in which the red shoulder is more prominent than the blue one, is shown in Fig. \ref{fig:triplehbeta} and corresponds to parameter values of $\xi_1 = 100 \ R_{g}$, $\xi_2 = 10000 \ R_{g} $, $q=1$, $ \sigma_{disk} = 240 \ km \ s^{-1}$ , $i=10\deg$ , $e=0.77$, and $\phi_0 = 120 \deg$. The $\sigma_{disk} = 240 \ km s^{-1}$ parameter is close to the resolution of the observation, and we note that the low inclination angle, consistent with an almost face-on disk, is consistent with the observed strong X-ray emission, and the presence of Fe II lines, according to TDE unified models (e.g., \citealp{dai2018, charalam2023}).

\section{Discussion}
\label{sec:discu}

We have presented multiwavelength time-evolving properties of the X-ray discovered transient J012026. Based on the nuclear position of the source, and the high X-ray luminosity in the 0.2-2.3 keV band ($L_x=3.1\times10^{43}$ ergs/s) alone, the source is most likely associated with an accretion event onto an SMBH. Furthermore, the soft X-ray spectrum, the multiwavelength flares (in the MIR and X-ray bands), the He\,II, and Bowen Fluorescence lines observed in the follow-up optical spectra, together with the absence of evidence of previous AGN activity point towards a TDE origin (see e.g. \citealp{vanvelzen20, Saxton2020, gezari2021tidal, jiang2021infrared, vanvelzen21echoes}). However, the erratic X-ray behavior and the relatively narrow profiles of Balmer lines seem to contradict this interpretation. This renders, effectively, J012026 a member of the agnostic Ambiguous Nuclear Transient classification (ANT, e.g.,\citealp{hinkle2022curious}). In the following section, we further discuss the phenomenological properties and the physical nature of J012026.

\subsection{J012026 is a BFF and ECLE}

J012026 shares several properties with the transient class of Bowen fluorescence flares. These events might appear at first glance similar to TDEs, especially due to the presence of transient He\,II and Bowen Fluorescence lines such as N\,III, which are not to be expected in AGN (\citealp{netzer1997}). However BFFs are different from standard TDEs as their overall photometric evolution is slower (a few years rather than a few months), and the transient emission lines are narrower (a few $10^3$ km s$^{-1}$ rather than a few $10^4$ km s$^{-1}$) and more persistent than in TDEs (e.g. \citealp{Trakhtenbrot2019bff,makrygianni2023bff}).

Indeed, the X-ray, optical, and MIR flares of J01206 evolved over a timescale of about 1000 days, on the longer end of typical optically selected TDEs (e.g., \citealp{vanvelzen20, Saxton2020, jiang2021infrared, vanvelzen21echoes}), although we note that MIR flares of similar profiles have been used to select infrared populations of TDEs (\citealp{masterson2024new}). 
Moreover, both Balmer and He\,II+N\,III lines detected in the spectra of J012026 have FWHM $<$ 3000 km s$^{-1}$ and persist for $>$ 600 days, compatible with typical BFF spectral properties. The Balmer lines, in particular, are detected up to 4 years after the initial discovery.

An additional feature that J012026 shares with BFFs is the presence of coronal lines. Indeed,
most sources of the same category were also found to emit strong and, in some cases, extreme CLs. This is the case for F01004-2237 (\citealp{F01004-2237}), AT 2021loi (\citealp{makrygianni2023bff}), AT 2019avd \citealp{at2019avd}), AT 2017bgt (\citealp{Trakhtenbrot2019bff}), AT 2022fxp \citealp{koljonen2024at2022fpx}), AT 2019aalc (\citealp{veres2024at2019aalc, Sniegowska2025}), and AT 2019pev (\citealp{frederick2021family}). As discussed in Section \ref{sec:optspec}, the coronal lines [Fe\,X] and [Fe\,XIV](+[Fe\,VII]) are strongly detected in the follow-up spectra of J012026. Interestingly, they have comparable intensity to [O\,III] 5007, making J012026 also part of the ECLE class. Following the discussion in \citealp{veres2024at2019aalc}, which also reports a nuclear transient associated with both the BFF and ECLE class, we confirm the connection between ECLEs and strong MIR flares. This supports the scenario proposed in \citealp{wang2012ecle}, and observationally supported for the first time by the work of \citealp{onori2022nuclear}, in which coronal lines arise due to dust-rich environments being sublimated by TDE-like flares and exposing iron atoms to the ionizing field.

\begin{figure}
    \centering
    \includegraphics[width=\columnwidth]{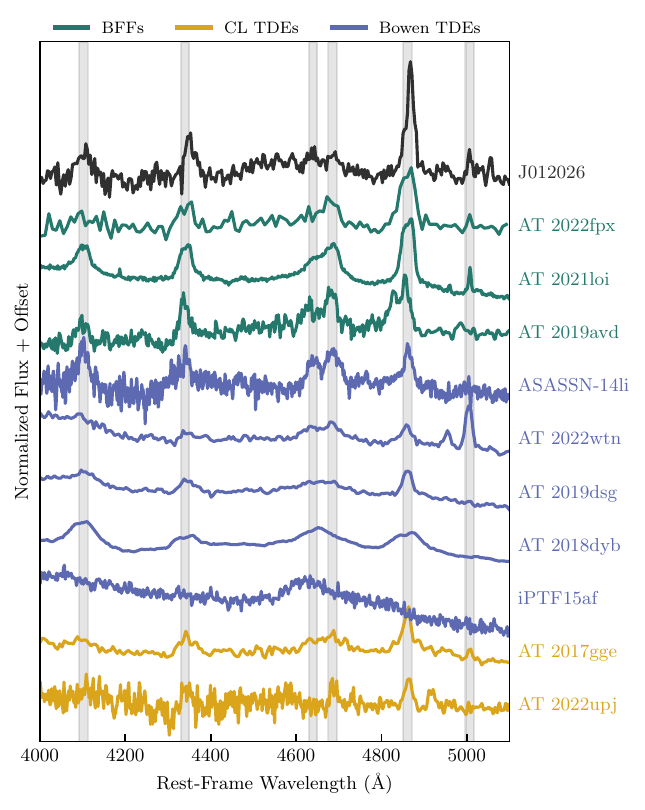}
    \caption{Comparison of spectrum 3 of J012026 with examples of the different transient classes discussed transients: BFFs (In green, AT 2021loi, \citealp{makrygianni2023bff}, AT 2019avd \citealp{malyali2023erasst,frederick2021family}, and AT 2022fpx \citealp{koljonen2024at2022fpx}), TDEs with Bowen features (in blue, ASASSN-14li \citealp{Holoien201614li}, AT 2022wtn \citealp{onori2025at2022wtn}, AT 2019dsg \citealp{vanvelzen2021seventeen}, AT 2018dyb \citealp{leloudas2019spectral}, and iPTF15af, {\citealp{Blagorodnova2019}}), and TDEs with coronal lines (in yellow, AT 2017gge,\citealp{onori2022nuclear}, and AT 2022upj,\citealp{Newsome2024mapping}). The grey shaded areas show, respectively from left to right, the wavelength corresponding to H$_\delta$, \hg, N\,III, He\,III, \hb and [O\,III]$\lambda$5007. The presented spectra are retrieved from the transient name server or from the ESO archive.)}
    \label{fig:spec_comparison}
\end{figure}

A summary and visual representation of the spectroscopic similarities and differences between J012026, TDEs, BFFs, and ECLEs, can be seen in Fig. \ref{fig:spec_comparison}. In the figure, the spectrum of J012026 is compared with the example spectra of the other discussed classes of transients. The similarities between J012026 with BFFs and CL TDEs appear stronger than with the typical Bowen TDE case (AT 2019dsg, AT 2018dyb, and iPTF15af, respectively \citealt{vanvelzen2021seventeen, leloudas2019spectral, Blagorodnova2019}). However, we note that for two Bowen TDEs (ASASSN-14li and AT 2022wtn, respectively \citealp{Holoien201614li, onori2025at2022wtn}), at least in one epoch, the emission line profile of the He\,II and N\,III complex appeared narrow and resolved, resembling that of BFFs, and possibly hinting at a connection between these classes.      

J012026, however, presents several unique features with respect to the aforementioned classes. Indeed, BFFs are characterized by late-time re-brightenings in their optical light curves (e.g., \citealp{koljonen2024at2022fpx}). In the case of J012026, we do not have evidence for such re-brightening in the optical lightcurve. However, we do observe such feature in the X-rays, both on time scales of a few hours and of a few months (Fig. \ref{fig:erolc}). In fact, the eRASS5 re-brightening brings the baseline flux back to its discovery eRASS2 level for at least one day. The separation between the peaks cannot be constrained accurately, but it is between 1 and 2 years. This is similar to the 1-year timescale observed in the optical lightcurves of AT 2022fpx and AT 2021loi, and possibly AT 2019pev (\citealp{koljonen2024at2022fpx,makrygianni2023bff,frederick2021family}), and to the X-ray behavior of AT 2019aalc (\citealp{veres2024at2019aalc, Sniegowska2025}). The similarity of the timescales could be connected to the inferred SMBH masses, which are of the order of $M_{SMBH} \sim 10^7 M_{\odot}$ for all of the mentioned sources. \citealp{koljonen2024at2022fpx} notes that the differences between the amplitude of the optical re-brightening in AT 2022fpx (lower) and AT 2021loi (higher) and an X-ray rebrightening being only confidently detected in AT 2022fpx could be ascribed to the presence of larger amounts of circumnuclear material reprocessing the primary accretion-powered X-ray emission into optical wavelengths. In a scenario in which BFFs are powered by TDEs reactivating dormant AGN, the unique long-term X-ray lightcurve of J012026 would fit in the picture as an event observed nearly face-on, where we expect the X-ray emission to be the strongest and to experience little reprocessing along the line of sight (\citealp{dai2018}). The orientation claim is supported by the presence of strong Fe\,II (\citealp{charalam2023}) and by the inclination inferred by modeling the disk lines in spectrum 5 (See section \ref{sec:tphb}). However, we note that the physical origin of the re-brightening has yet to be clarified.

In section \ref{sec:optspec}, we have highlighted how the Balmer and He\,I lines are systematically shifted towards the red, with respect to the higher ionization lines. We suggest that this evidence points towards the contribution to the lines coming from different kinematical components of the transient, as also argued for some TDEs, such as AT 2022wtn (\citealp{onori2025at2022wtn}). In fact, He\,II, N\,III and the CLs are likely to originate around the X-ray emitting accretion disk due to the required high ionization radiation field. Instead, the Balmer lines seem to receive a significant contribution from an asymmetrical, non-spherical outflow, which, due to our line of sight, makes them appear receding. One possible origin of this receding material, in the case of TDEs, is a collision-induced outflow (CIO, \citealp{lubonnerot, jiang2016A}), happening at the self-intersection point, which is naturally non-spherical. In this context, the Balmer lines also appear reddened in the latest spectrum, either due to self-absorption or because of the cooling-down of the nuclear material after the event is terminated. 

We note, however, that in the rejuvenated SMBH interpretation of BFFs, the Balmer, He\,II and N\,III lines are all produced in a pre-existing BLR-like structure. The observed dynamical decoupling of these components in J012026, shown by both the width and different shifts of the lines, adds complexity to the physical interpretation of BFFs. This is also the case for the puzzling appearance of the triple-peaked \hb in spectrum 5, although we suggest that a local obscuration event could temporarily suppress the main emission component and expose the weaker, underlying disk emission in that epoch.

\subsection{On the short-time X-ray variability}

As mentioned in section \ref{sec:discovery}, J012026 caught our attention not only as a TDE, but also because it triggered the QPE detection pipeline developed in \citealp{Arcodia2024cosmic}, which is based on the detection of significant eROday variability. Indeed, the eRASS3 and eRASS5 lightcurves show variability on eRO-Day timescales, and the second and third Swift XRT observations have also detected high flux states comparable to the total eRASS2 or eRASS3 and eRASS5 flare fluxes. The Swift XRT snapshots are of the order of 1\,ks, therefore the flaring timescales are possibly similar to what is observed in the eROSITA lightcurves (40s). Additionally, we see a tentative harder-when-brighter behavior, consistent with the QPE spectral evolution as shown in Fig. \ref{fig:hwb}, and discussed in Section \ref{sec:x-ray}.

The two flares in eRASS5 are separated by 16 hours each. Our longest observation with XMM-Newton covers a baseline of 12 hours, making it possible that, if the flares are due to QPEs with a period of 16 hours, our observation fell right in the quiescence period and missed any new flare. The latest Swift XRT observations, however, which last 400s, with a cadence of 1-2 hours over a total baseline of three days, also resulted in non-detections.
Therefore, if the source is indeed emitting QPEs: 1) the amplitude of the peaks is either below the Swift-XRT upper limits, or 2) the period is longer than expected, or 3) very irregular (see e.g. ZTF19acnskyy, \citealp{hernandezgarcia2025ansky}), or 4) the source is experiencing a quiet-phase, similarly to what has been observed for GSN069 (\citealp{miniutti2019nine}). While all of these possibilities are feasible, and we encourage further X-ray follow-up of this source, we note that the evidence pointing towards the variability of J012026 being due to QPEs is yet tentative in spite of our dedicated follow-up programs. Therefore, we explore other models that could explain the short-timescale flares. 

One possible scenario explaining the X-ray variability is supported by the disk orientation suggested by the triple-peaked \hb and by the X-ray brightness of the event. In such a configuration, the line of sight could intercept the inner regions of the accretion disk with little to no reprocessing material to smooth out any accretion flow-driven variability. Indeed, accretion flows are known to be variable on timescales spanning several orders of magnitude (e.g. \citealp{Ulrich1997,czerny2004role}). 
In the case of J012026, the observed shortest timescale of X-ray variability is 16 hours eRASS4, which is fully consistent with the thermal timescale at 3-10 $R_g$ for a $10^7 M_{\odot}$ SMBH, assuming an $\alpha-$disk with $\alpha=0.1$ (\citealp{czerny2004role}). Therefore, thermal fluctuations in the innermost regions of the accretion disk could potentially explain the erratic X-ray behavior, although the amplitude of the variability is quite extreme. 

Alternatively, another scenario that could give rise to short-time X-ray variability can be connected to the high-density environment in which the TDE is expected to be taking place, as suggested by the MIR flare, the CLs, and the BFF classification. The simulations of \citealp{ryu2024} of TDEs in AGN have revealed that a star's passage inside a disk will perturb it and cause instabilities. Even though their analysis focused on the TDE debris, it was observed that these perturbations can cause the disk gas to rain down onto the SMBH. We speculate that these episodic accretion events could superimpose on the TDE light curve and cause short-time X-ray variability on free-fall timescales. In this scenario, in order for the flares to be present in eRASS3, SWIFT-XRT3, SWIFT-XRT4 and eRASS5, perturbations to the disk need to reoccur every $\sim5$ months. This would in principle be feasible if the disrupted star was on a bound orbit, and the TDE was not full, but rather partial (pTDE, e.g. \citealp{liu2023deciphering,liu2024rapid}). We note, however, that our observations generally do not show any evidence supporting a pTDE scenario, although currently, no numerical work on the emission signatures of pTDE in high-density environments/AGN disks exists, to our knowledge. Overall, the extreme multi-scale X-ray variability of J012026 remains puzzling, and further theoretical work is needed. This is especially crucial if the origin of the X-ray rebrightenings in J012026 is connected to the optical rebrightenings in other members of the BFF class.

\section{Summary and Conclusions}
\label{sec:sum}
We have presented the multiwavelength observation of the unique nuclear transient J012026. We summarize the main results in the following bullet points:

\begin{itemize}
    \item J012026 was discovered as a bright $F_{0.2-2.3} = 1.14\times 10^{-12}$\ erg/s/cm$^2$ and soft ($\Gamma=4.3$) nuclear X-ray transient in eRASS2. Subsequent eRASS scans and follow-up observations confirmed the transient nature of the source, and revealed X-ray variability both on hour and month timescales. 

    \item The multiwavelength light curves from ATLAS and WISE revealed the presence of long-lived flares ($\sim2$ years). The MIR flare has an amplitude of 0.5-0.7 mag, while the optical flare is less constrained and with evidence for a flux increase $>3$ year prior to the X-ray discovery.

    \item Follow-up optical spectroscopy revealed a richness of transient features, not detected in the archival spectrum, namely 1) redshifted Balmer lines with FWHM$\sim$1500 km s$^{-1}$, lasting for over 4 years, 2) He\,II and the Bowen N\,III line with FWHMs$\sim$2000 km s$^{-1}$, 4) Strong Fe\,II, 5) high ionization CLs  ([Fe\,X],[Fe\,XIV] and possibly [Fe\,VII]), and 6) a triple-peaked \hb line in one spectrum, suggesting an origin from an elliptical disk-like structure.    

    \item The optical spectral features make this object part of the BFF and ECLE class, supported by the long duration of the event, and the strong MIR flare. Optical re-brightenings are common among BFFs, yet J012026 shows this feature only in the X-rays, unlike any other source of this class. The re-brightening timescales and black hole masses are also compatible with events AT 2021loi and AT 2022fpx. In the interpretation in which BFFs are TDEs rejuvenating AGN, J012026 would represent a face-on event based on the X-ray to optical properties and optical spectral features. 

    \item We discussed possible models to explain the short- and long-term X-ray variability, however, leaving the question still unresolved. We stress the need for more theoretical predictions of the X-ray properties of full and partial TDEs in high-density environments (such as AGN disks).

\end{itemize}

J012026, overall, shows a complex interplay of features, bridging the classification of nuclear transients of TDEs, ECLEs, BFFs, and ANTs. The richness in multiwavelength features points towards an ambiguous origin of the transient, and more theoretical and observational work is encouraged to narrow down the physical origin of analogous sources. We highlight that, without the early X-ray detection, this source would not have been selected as optically variable. This is testimony to the importance of X-ray time-domain surveys, such as eROSITA and Einstein Probe, to gain a full picture of nuclear transients. 

\begin{acknowledgements}
PB is grateful for the insightful discussions with Megan Newsome, Paola Martire, and Fabian Balzer about the work. This work is based on data from eROSITA, the soft X-ray instrument aboard SRG, a joint Russian-German science mission supported
by the Russian Space Agency (Roskosmos), in the interests of the
Russian Academy of Sciences represented by its Space Research Institute (IKI), and the Deutsches Zentrum für Luft- und Raumfahrt
(DLR). The SRG spacecraft was built by Lavochkin Association
(NPOL) and its subcontractors, and is operated by NPOL with support from the Max Planck Institute for Extraterrestrial Physics (MPE).
The development and construction of the eROSITA X-ray instrument was led by MPE, with contributions from the Dr. Karl Remeis Observatory Bamberg \& ECAP (FAU Erlangen-Nuernberg),
the University of Hamburg Observatory, the Leibniz Institute for
Astrophysics Potsdam (AIP), and the Institute for Astronomy and
Astrophysics of the University of Tübingen, with the support of DLR
and the Max Planck Society. The Argelander Institute for Astronomy
of the University of Bonn and the Ludwig Maximilians Universität
Munich also participated in the science preparation for eROSITA.
The eROSITA data shown here were processed using the \texttt{eSASS}
software system developed by the German eROSITA consortium.

Some of the observations reported in this paper were obtained with the Southern African Large Telescope (SALT), as part of the Large Science Programme on transients 2018-2-LSP-001 (PI: Buckley).

This work is based on data obtained with
the Einstein Probe, a space mission supported by the Strategic Priority Program
on Space Science of the Chinese Academy of Sciences, in collaboration with
ESA, MPE and CNES (Grant No. XDA15310000, No. XDA15052100). 

R.A. was supported by NASA through the NASA Hubble Fellowship grant \#HST-HF2-51499.001-A awarded by the Space Telescope Science Institute, which is operated by the Association of Universities for Research in Astronomy, Incorporated, under NASA contract NAS5-26555. M.K. is supported by DLR grant FKZ 50 OR 2307.

This work was supported by the Australian government through the Australian Research Council’s Discovery Projects funding scheme (DP200102471)

The Australia Telescope Compact Array is part of the Australia Telescope National Facility (grid.421683.a) which is funded by the Australian Government for operation as a National Facility managed by CSIRO.

This work has made use of data from the Asteroid Terrestrial-impact Last Alert System (ATLAS) project. The Asteroid Terrestrial-impact Last Alert System (ATLAS) project is primarily funded to search for near earth asteroids through NASA grants NN12AR55G, 80NSSC18K0284, and 80NSSC18K1575; byproducts of the NEO search include images and catalogs from the survey area. This work was partially funded by Kepler/K2 grant J1944/80NSSC19K0112 and HST GO-15889, and STFC grants ST/T000198/1 and ST/S006109/1. The ATLAS science products have been made possible through the contributions of the University of Hawaii Institute for Astronomy, the Queen’s University Belfast, the Space Telescope Science Institute, the South African Astronomical Observatory, and The Millennium Institute of Astrophysics (MAS), Chile.

\end{acknowledgements}
\bibliographystyle{aa} 
\bibliography{J012026.bib}

\begin{appendix}
\section{Supplementary material on X-ray observations}

\subsection{Contamination from nearby AGN WISEA J012025-292637}
\label{sec:qso}

\begin{figure}[!h]
    \centering
    \includegraphics[width=0.7\columnwidth]{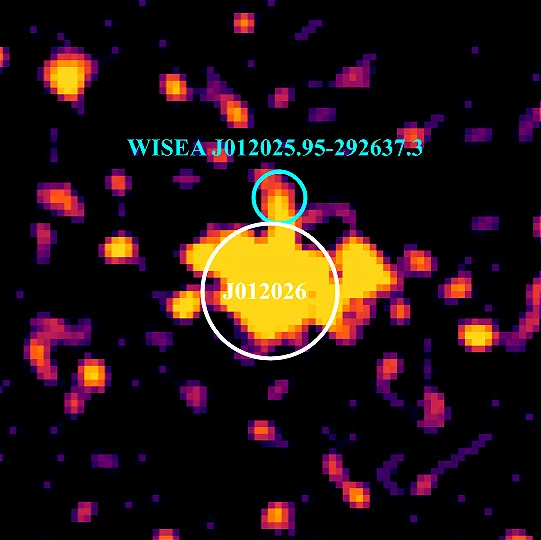}
    \caption{ eRASS:5 image of J012026 and nearby AGN WISEA J012025-292637. The image had been smoothed for visualization purposes. The white circle represent the extraction region of J012026 used in the procedure to estimate the contamination, while the cyan circle is the region used for WISEA J012025-292637.}
    \label{fig:doubleextract}
\end{figure}

\begin{figure*}[!h]
    \centering
    \includegraphics[width=\textwidth]{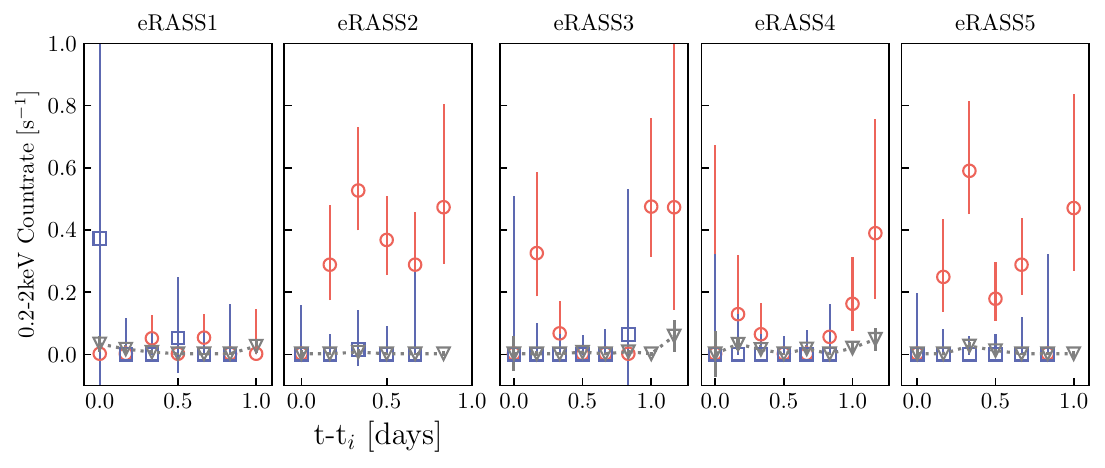}
    \caption{eROSITA eRO-day 0.2-2.3 keV band lightcurve of J012026 (red) and WISEA J012025-292637 (blue) extracted respectively with regions of 15" and 40" in radius. The grey triangles correspond to the background countrate}
    \label{fig:erolcqso}
\end{figure*}

\begin{figure*}[!h]
    \centering
    \includegraphics[width=\textwidth]{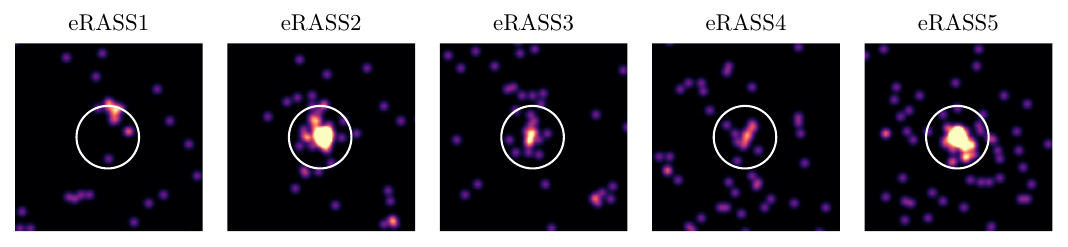}
    \caption{eROSITA images of J012026. The images have been smoothed for visualization purposes, and the white circle represents a region of 60" in radius.}
    \label{fig:cutouts}
\end{figure*}

To estimate the possible contamination of WISEA J012025-292637, we extracted light curves at the locations of J012026 and WISEA J012025-292637, in addition to an annulus background region of inner and outer radii of 115" and 160".
We used a circular region for J012026 with a radius of 40", corresponding to $\sim80\%$ of the encircled energy fraction (EEF). For WISEA J012025-292637, we used a region with a radius of 15" , which corresponds to $\sim50\%$ of the EEF. 
We chose such radii in order to avoid overlapping between the two regions, while still being able to estimate the total number of counts, through our knowledge of the eROSITA PSF. The extraction regions are shown in Fig. \ref{fig:doubleextract} over the stacked eRASS:5 image. 
We then computed the fraction $f$ of the remaining 50\% photons coming from  WISEA J012025-292637 that would fall in the extraction region of J012026, by measuring the angle between the lines tangent to such extraction region, passing through the WISEA J012025-292637 position.  

In Fig. \ref{fig:erolcqso}, we plot the lightcurves of J012026 and the maximum expected contamination by WISEA J012025-292637, estimated from the 50\% extraction region and corrected by the factor $f$, respectively in red circles and blue squares. We conclude that WISEA J012025-292637 was bright only in eRASS1, as also shown by the cutouts in Fig. \ref{fig:cutouts}, and did not contaminate the remaining observations. Therefore we use pipeline products.

\subsection{X-ray spectra and analysis of the harder-when-brighter behavior}
\label{xrayspec}

We show the eRASS2-5, SWIFT-XRT3 and SWIFT-XRT4 spectra in Fig. \ref{fig:xrayspeca}. The figures show both the source and background data and models, both for the diskbb and power-law fits. 

As discussed in section \ref{sec:x-ray}, the results of the X-ray spectral fits suggest a general anti-correlation between the X-ray flux and power-law photon index $\Gamma$. To quantitatively test this hypothesis, we performed a Monte Carlo evaluation of the Pearson correlation coefficient $r$, uniformly sampling our dataset accounting for uncertainties in both fluxes and photon indices. The resulting distribution of $r$ is shown in Figure~\ref{fig:MC}. The median value of $r = -0.56$ suggests a possible anti-correlation, however, the 95\% confidence interval of the distribution includes zero, indicating that the observed trend is not statistically significant. 

\begin{figure*}[t]
    \centering
    \includegraphics{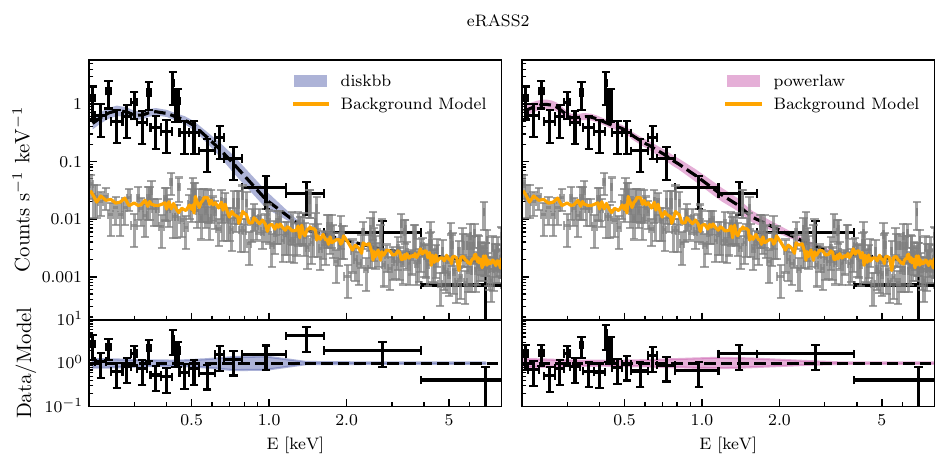}
    \includegraphics{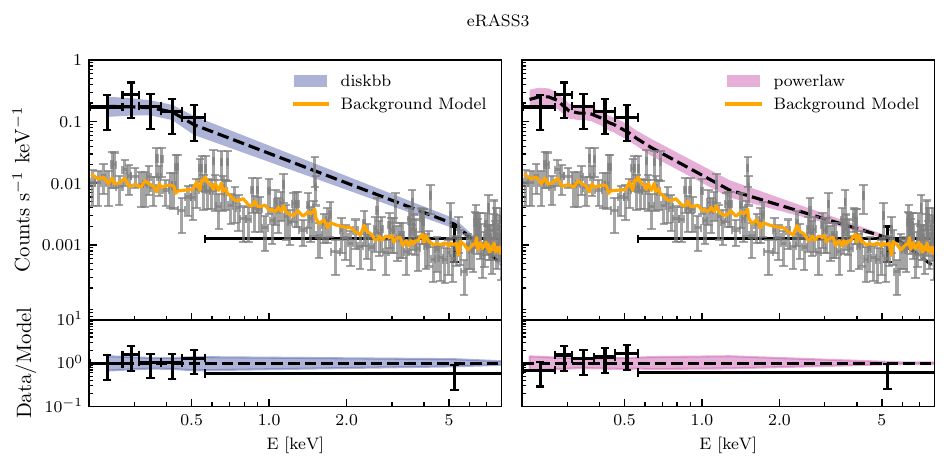}
    \includegraphics{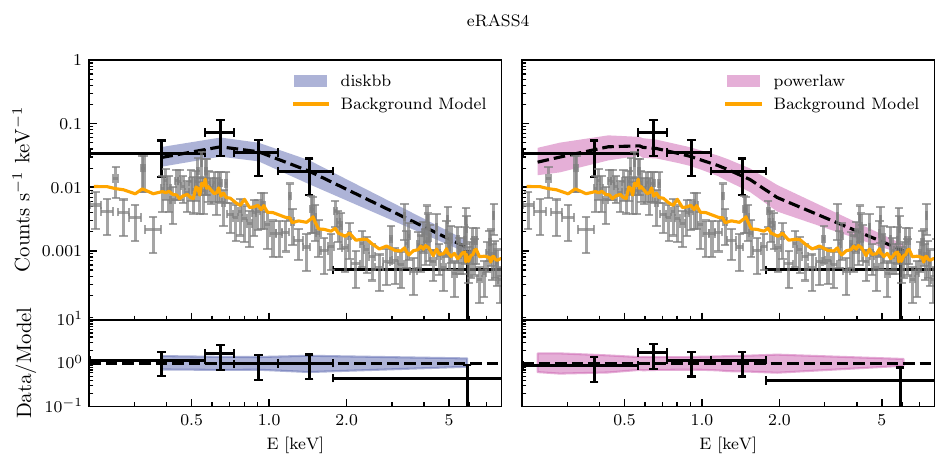}
    \caption{The X-ray spectra of J012026. The observation name is reported in the title. Both left and right panels show, respectively in black and grey markers, the source and background counts, which have rebinned for visualization purposes. The left panels show the diskbb fit (black dashed line) with 1$\sigma$ uncertainties (blue band), while the right panels show the same results for the power-law model (black dashed line and pink band). In both left and right panels, the orange line shows the background model.}
    \label{fig:xrayspeca}
\end{figure*}
\begin{figure*}[t]
    \centering
    \ContinuedFloat
    \includegraphics{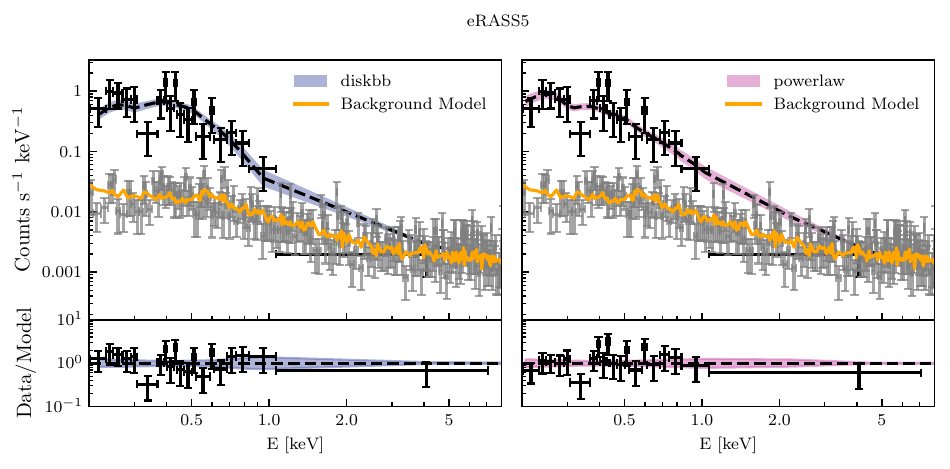}
    \includegraphics{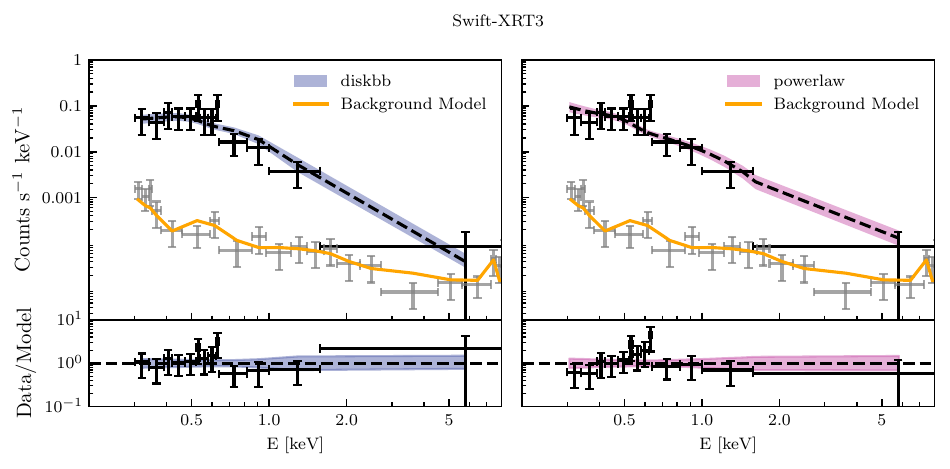}
    \includegraphics{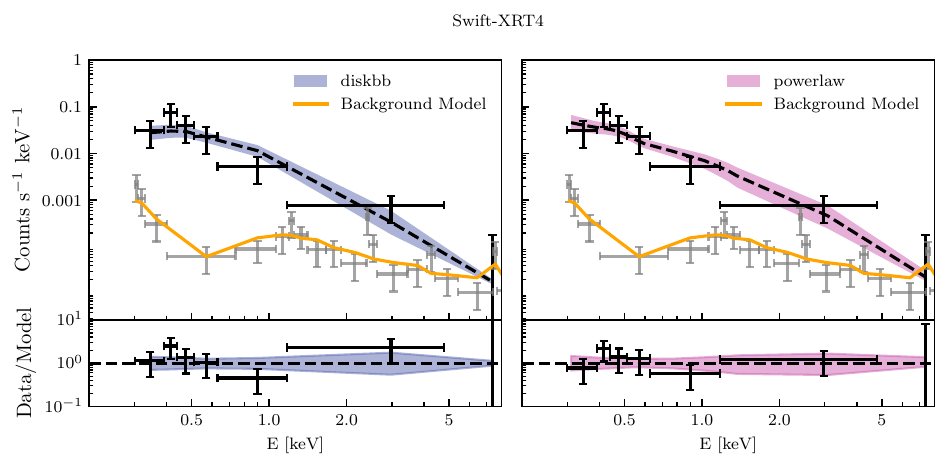}
    \caption{Continued from Fig. \ref{fig:xrayspeca}}
    \label{fig:xrayspecb}
\end{figure*}
\begin{figure*}[t]
    \ContinuedFloat
    \centering
    \includegraphics{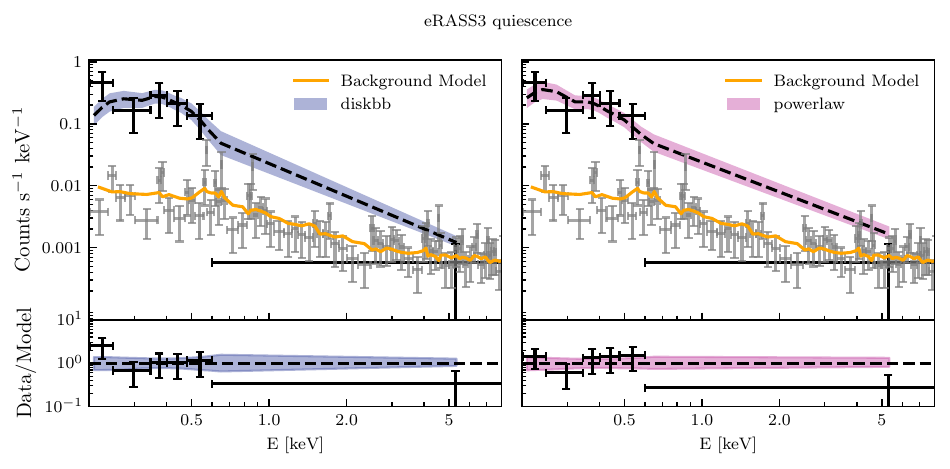}
    \includegraphics{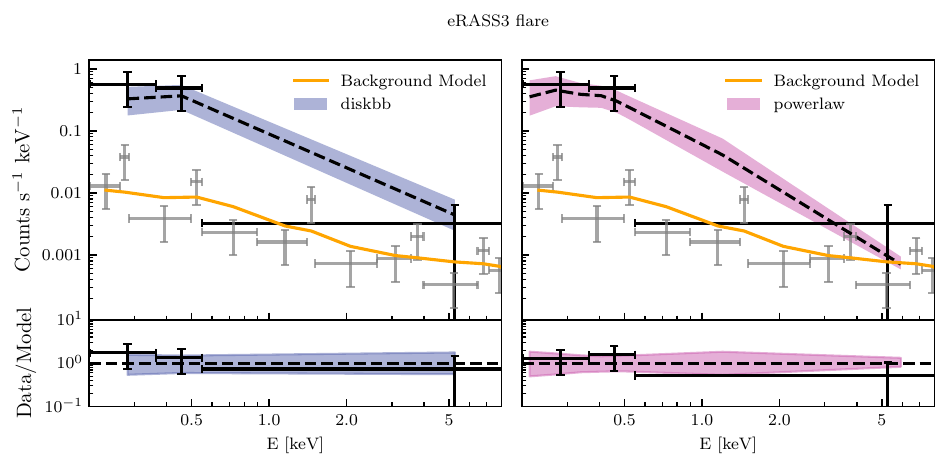}
    \caption{Continued from Fig. \ref{fig:xrayspeca}}
\end{figure*}
\begin{figure*}[t]
    \ContinuedFloat
    \centering
    \includegraphics{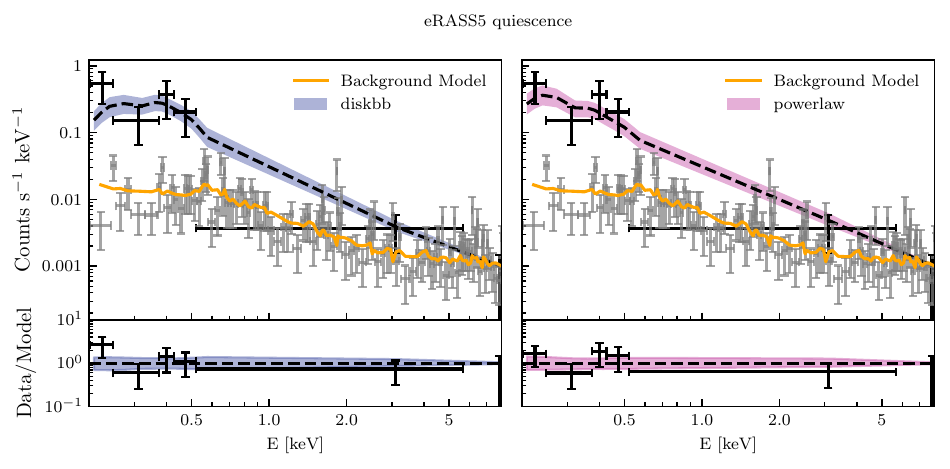}
    \includegraphics{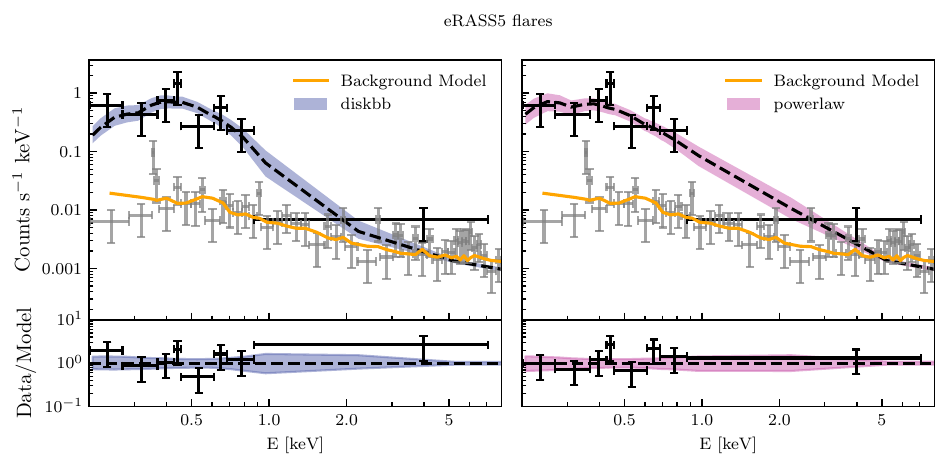}
    \caption{Continued from Fig. \ref{fig:xrayspeca}}
\end{figure*}

 \begin{figure}
     \centering
     \includegraphics[width=1\columnwidth]{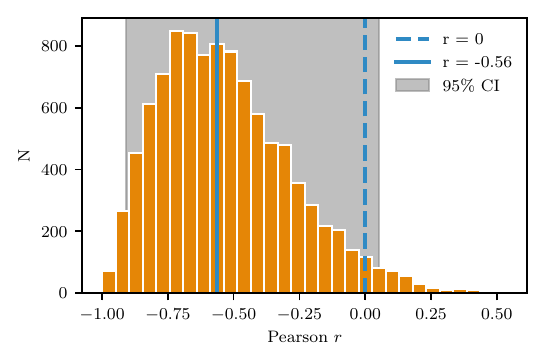}
     \caption{Distribution of the Pearson correlation coefficient $r$. The continuous vertical line indicate the median value of $r=-0.56$, while the dashed line shows $r=0$. The grey shaded area indicates the 95\% confidence interval.}  
     \label{fig:MC}
 \end{figure}

\section{Radio properties of J012026}
\label{app:radio}
In order to search for any transient radio emission associated with the multiwavelength outburst, we observed the coordinates of J012026 with the Australia Telescope Compact Array (ATCA) on three occasions in October 2022 (ATCA program C3513), June 2023 (ATCA program C3513), and March 2024 (ATCA program C3586). In the first and third observations we observed with the dual 5.5 and 9\,GHz receiver with 2\,GHz of bandwidth split into 2048 spectral channels at each central frequency. In the second observation we observed with the dual 5.5 and 9\,GHz receiver as well as the 2.1\,GHz receiver with 2\,GHz of bandwidth split into 2048 spectral channels. All observations were taken in the extended 6\,km configuration. The observations in October 2022 and March 2024 were approximately 4\,hr long and the observation in June 2023 was approximately 6\,hr long. In all observations, PKS 1934--638 was used for flux and bandpass calibration and PKS 0142-278 for phase calibration. All data were reduced using standard procedures in the Common Software Astronomy Application \citep[CASA][]{CASA2022}. Images of the target field were created using the CASA task \texttt{tclean} and target flux density extracted by fitting a Guassian the size of the synthesized beam using the CASA task \texttt{imfit}. In all observations except the 5.5\,GHz observation on 2022-10-03 and 2024-03-25 no radio source was detected at the coordinates of J012026, with a 3$\sigma$ upper limit extracted as 3 times the image rms. In the observations at 5.5\,GHz on 2022-10-03 and 2024-03-25, a marginal point source was detected at the location of J012026 with a flux density of $55\pm9$ and $37\pm6$\,uJy, i.e. a 6$\sigma$ detection in each observation. The source was not detected in the 5.5\,GHz observation on 2023-06-23. We report the flux density measurements and upper limits in Table \ref{tab:atca_obs}. 

The two marginal radio detections of J012026 at 5.5\,GHz may indicate variable jet activity related to the transient activity in the galaxy, although, without spectral information, it is difficult to determine the nature of the radio emission. However, the detections are marginal, and may also be spurious noise peaks in the images.

\begin{table}
    \centering
    \begin{tabular}{ccc}
         Date & Frequency (GHz) & Flux Density (uJy)  \\
         \hline
        2022-10-03 & 5.5 & 55$\pm$9  \\
        2022-10-03 & 9 & $<44$ \\
        2023-06-23 & 2.1 & $<144$\\
        2023-06-23 & 5.5 & $<43$ \\
        2023-06-23 & 9 & $<41$\\
        2024-03-25 & 5.5 & 37$\pm$6\\
        2024-03-25 & 9 & $<24$\\
        \hline
    \end{tabular}
    \caption{The ATCA radio observations of J012026. All upper limits reported are the 3$\sigma$ upper limit.}
    \label{tab:atca_obs}
\end{table}

\section{Supplementary material on optical spectroscopy.}
\label{app:opt}
\subsection{The archival spectrum of J012026}
\begin{figure*}[t]
    \centering
    \includegraphics{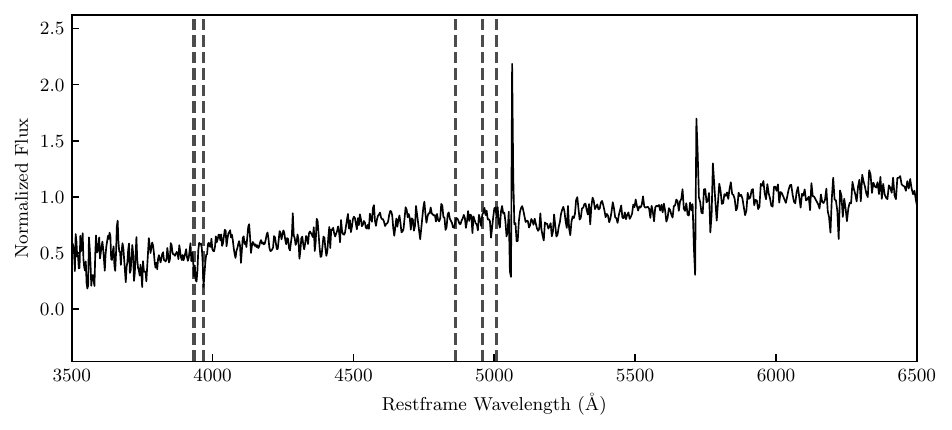}
    \caption{Archival 2dF spectrum of the nucleous of J012026. The black vertical lines show the resftrame wavelengths of the Ca H \& K doublet, \hb, and the [O\,III] doublet. The spikes slightly reward of the [O\,III] doublet and around $\sim5700 \AA$ are likely artifacts.}
    \label{fig:archspec}
\end{figure*}

We present the archival spectrum of J012026 in Fig. \ref{fig:archspec}, as retrieved from the NASA/IPAC Extragalactic Database. The spectrum was obtained by the 2dF spectrograph on the Anglo-Australian Telescope (AAT), and has a resolution of $\Delta \lambda \sim 9 \AA$. No evident features are present, with the exception of the Ca H \& K doublet.

\subsection{Spectral reduction}

NTT: The source was observed with the ESO Faint Object
Spectrograph and Camera v.2 (EFOSC2; \citealt{1984Msngr..38....9B}) mounted on the ESO New Technology Telescope (NTT) in La Silla, Chile (proposal ID: 108.225J.001, PI Grotova). The grism 13 and the 1.2\arcsec slit were used for the observation, providing a wavelength range of $\mathrm{3685-9315\AA}$ with a dispersion of $\mathrm{2.77\AA/pixel}$. The observations were performed with an airmass of 1.3. The data reduction and calibration were performed with the esoreflex pipeline \citep[][v2.11.5]{2013A&A...559A..96F}. For wavelength and flux calibration we used the He+Ar arcs and a standard star observed with the same grism and slit, oriented along the parallactic angle.

MagE: We observed J012026 with the Magellan Echellette (MagE) Spectrograph \citep{Marshall2008} mounted on 6.5m Magellan/Baade Telescope as Las Campanas Observatory on 16 December 2023. The total observations duration was 1 hour, with an airmass of 1.06, and the 0.5\arcsec slit was used. We reduced the data with \texttt{pypeit} \citep{Prochaska2020} and followed standard data reduction procedures, including wavelength calibration with ThAr lamps and flux calibration with observations of a nearby standard star on the same night. 

WiFeS: We obtained a spectrum of J012026 with the Wide Field Spectrograph \citep[WiFeS][]{dop2010} mounted on the 2.3m ANU telescope at Siding Spring Observatory. We used the R3000 and B3000 gratings providing a resolution of $R = 3000$ and a spectral coverage of 3500 to 7000 \AA. The observations were performed with an airmass of 1.02. The data were reduced using the PyWiFeS reduction pipeline \citep{chi2014}, which produces three-dimensional data cubes. The spectra in these cubes were bias subtracted, flat-fielded, wavelength (with a NeAr lamp) and flux calibrated (using the flux standard EG 131). Finally, the spectra were extracted from the slitlets that captured J012026 using the IRAF task APALL \cite{tod1986} which allowed for background subtraction.

SALT: We obtained SALT spectra on the nights of 2020-07-18 and 2021-07-17 using the Robert Stobie Spectrograph (RSS, \citealp{burgh2003prime}) on the South African Large Telescope (SALT, \citealp{buckley2006completion}). For both observations, the airmass was 1.18. We used the PG0900 grating with a 1.25 arcsec slit. The wavelength range is from 4060 to 7110~\AA with a typical resolution of R $\sim$1200. Observations were reduced using a custom pipeline based on the PySALT package (\citealp{crawford2010pysalt}), which accounts for basic CCD characteristics (e.g., cross-talk, bias and gain correction), removal of cosmic rays, wavelength calibration, and relative flux calibration. Standard IRAF/Pyraf routines were used to accurately account for sky background removal. Absolute flux calibration with SALT is difficult because of the telescope design, which has a moving, field-dependent and under-filled entrance pupil. Observations of spectrophotometric flux standards can, at best, only provide relative flux calibration (see, e.g., \citealp{buckley2018comparison}), which mostly accounts for the low-frequency telescope and instrument sensitivity changes as a function of wavelength.

IMACS and LDSS3-C: For the Baade IMACS observation, we used the longslit mode with Grism 300-17.5 which covered a range of 4254-9454 \AA\ with a dispersion of 1.27 \AA/px. In combination with a slit of 0.7$"$ the instrumental resolution was about 4.7 \AA. The total exposure time was 900 sec, the seeing was approximately around 0.8$"$, and the airmass 1.4.
For the Clay LDSS3-C  observation, we used the VPH-All grating with a 1$"$ slit. The grating covered a range of 3565-10608 \AA\ with a dispersion of 1.99 \AA/px and a resolution of about 8.3 \AA. A total exposure time of 2100 sec was applied under a seeing of about 1.3 $"$, with an airmass of 1.54.
In both cases, spectra were reduced with IRAF following the usual procedure of overscan and flat-field correction, wavelength calibration by means of a He-Ne-Ar lamp, and flux calibration through the observation of a spectrophotometric standard star. LTT1788 was observed with IMACS and the same slit used for the target, while LTT2415 was observed with LDSS3-C and a slit of 6$"$. Finally, the night-sky contribution was subtracted fitting each spectrum across the slit and then, the single exposures were averaged together.

\subsection{Spectral fitting}

\begin{figure*}[!t]
    \centering
    \includegraphics[width=\columnwidth]{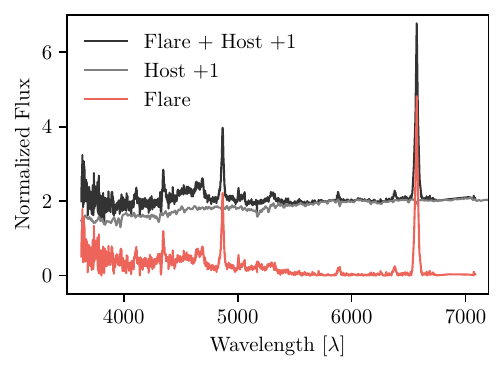}
    \includegraphics[width=\columnwidth]{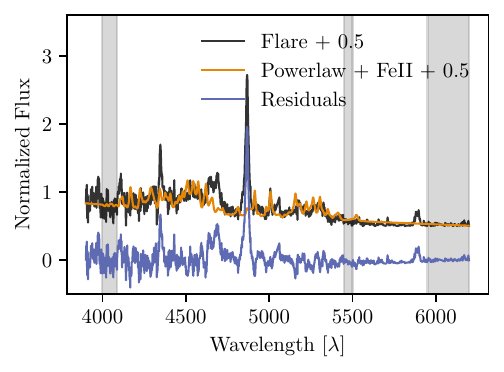}
    \caption{\textit{Left)} Example of the host subtraction procedure on spectrum 3. \textit{Right)} Example of the power-law and Fe II fitting and subtraction procedure on the red residuals of the left panel. The grey-shaded regions indicate where the power-law is fit.}
    \label{fig:continuumremoval}
\end{figure*}

\begin{figure*}
    \includegraphics[width=\textwidth]{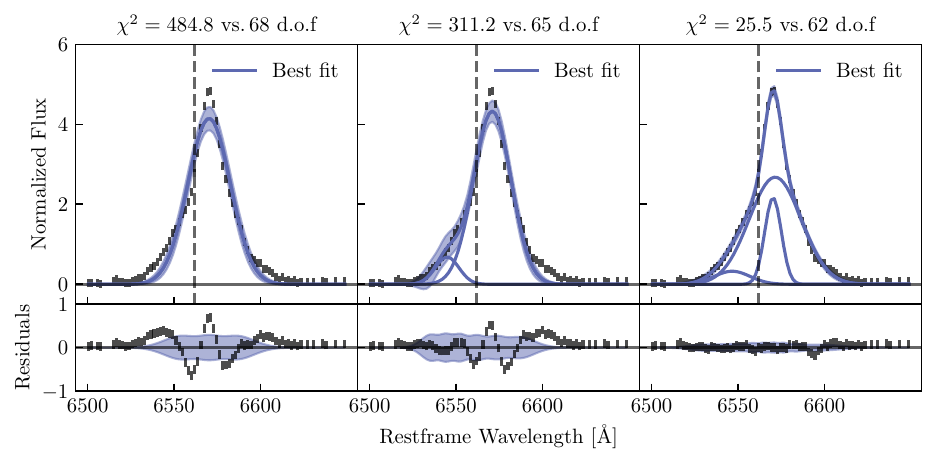}
    \includegraphics[width=\textwidth]{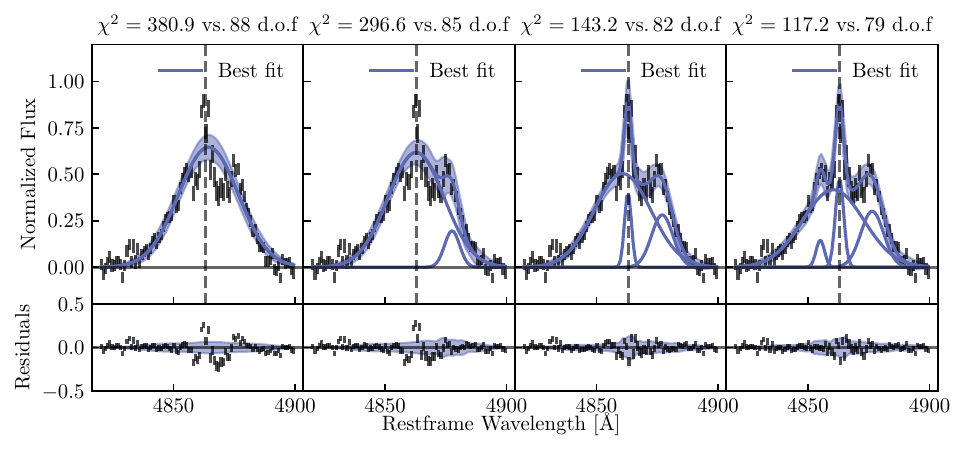}
    \caption{Emission line fitting procedure examples. \textit{Top:} example on the \ha line in spectrum 3. The three panels show respectivelly the best fit model obtained with one, two or three gaussians, and their respective residuals. The spectrum is represented by the black markers while the blue lines show the individual Gaussian components. The semi-transparent band shows the 3$\sigma$ uncertainty of the total model. The vertical dashed line indicates the restframe centroid of \ha. \textit{Bottom:} example on the triple-peaked \hb line in spectrum 5. The vertical dashed line indicates the restframe \hb centroid.}
    \label{fig:linefitting}
\end{figure*}

We show an example of the continuum subtraction procedure described in the main text in Fig. \ref{fig:continuumremoval}. 
In Fig. \ref{fig:linefitting}, we show two example of the line-fitting procedure described in the text. We start by fitting one Gaussian model component, and add up to three more, as long as the $\chi^2$ significantly improves. In the top panel of the figure, we show an example where a total of three Gaussians were needed to properly reproduce the \ha line (spectrum 3). In the lower panel, we show the case of the triple-peaked \hb line in spectrum 5, where four Gaussians are necessary for reproducing the emission complex. The emission line parameters obtained through this procedure for all spectra are shown in Table \ref{tab:alllines}. Given that all spectra were obtained with different instruments and set ups, we do not report the absolute line fluxes. Instead, we report relative fluxes to \hb, the only line detected in all spectra. 

\begin{sidewaystable*}
\centering
\caption{Line measurements for each of the 7 follow-up spectra. Fluxes are reported relative to \hb. All values are in rest-frame.}
\label{tab:alllines}
\begin{tabular}{ccccccccc}
\hline
\hline
ID & Parameters & H$\gamma$ & [N\,\textsc{iii}] & He\,\textsc{ii} & H$\beta$ & He\,\textsc{i} & [Fe\,\textsc{x}] & H$\alpha$ \\
\hline
1 & Centroid (\AA)   & 4344.78 $\pm$ 1.00 & - & 4683.65 $\pm$ 0.91 & 4866.10 $\pm$ 0.30 & - & 6378.00 $\pm$ 1.00 & -  \\
  & FWHM (km s$^{-1}$) & 2074.9 $\pm$ 170.0 & - & 2409.6 $\pm$ 143.0 & 1585.0 $\pm$ 44.0 & - & 899.0 $\pm$ 100.0 & -  \\
  & Relative flux    & 0.32 $\pm$ 0.13 & - & 0.38 $\pm$ 0.04 & 1.00 & - & 0.06 $\pm$ 0.01 & - \\
\hline
2 & Centroid (\AA)   & 4348.21 $\pm$ 1.39 & - & - & 4868.38 $\pm$ 0.50 & - & - & - \\
  & FWHM (km s$^{-1}$) & 1135.1 $\pm$ 224.0 & - & - & 1385.8 $\pm$ 73.0 & - & - & - \\
  & Relative flux    & 0.35 $\pm$ 0.10 & - & - & 1.00 & - & - & - \\
\hline
3 & Centroid (\AA)   & 4349.07 $\pm$ 0.90 & 4640.70 $\pm$ 1.50 & 4690.03 $\pm$ 2.80 & 4867.62 $\pm$ 0.43 & 5883.93 $\pm$ 1.53 & 6370.00 $\pm$ 2.00 & 6570.65 $\pm$ 0.25 \\
  & FWHM (km s$^{-1}$) & 1389.9 $\pm$ 146.0 & 1410.0 $\pm$ 228.0 & 2601.0 $\pm$ 572.0 & 1417.8 $\pm$ 50.0 & 1184.0 $\pm$ 184.0 & 1026.0 $\pm$ 185.0 & 1718.9 $\pm$ 35.0 \\
  & Relative flux    & 0.31 $\pm$ 0.13 & 0.15 $\pm$ 0.02 & 0.26 $\pm$ 0.05 & 1.00 & 0.11 $\pm$ 0.02 & 0.07 $\pm$ 0.02 & 3.53 $\pm$ 0.39 \\
\hline
4 & Centroid (\AA)   & 4348.95 $\pm$ 0.80 & 4641.49 $\pm$ 0.93 & 4684.93 $\pm$ 1.10 & 4866.52 $\pm$ 0.30 & 5888.56 $\pm$ 0.70 & 6376.50 $\pm$ 1.00 & 6570.23 $\pm$ 0.36 \\
  & FWHM (km s$^{-1}$) & 1263.1 $\pm$ 119.0 & 1132.0 $\pm$ 132.0 & 2665.0 $\pm$ 193.0 & 1490.7 $\pm$ 47.0 & 1247.0 $\pm$ 85.0 & 1424.0 $\pm$ 100.0 & 1307.2 $\pm$ 39.0 \\
  & Relative flux    & 0.24 $\pm$ 0.04 & 0.13 $\pm$ 0.03 & 0.42 $\pm$ 0.05 & 1.00 & 0.10 $\pm$ 0.01 & 0.15 $\pm$ 0.02 & 2.18 $\pm$ 0.67 \\
\hline
5 & Centroid (\AA)   & 4345.36 $\pm$ 0.58 & - & 4683.52 $\pm$ 0.60 & 4864.10 $\pm$ 0.50 & 5883.77 $\pm$ 1.10 & 6374.33 $\pm$ 1.00 & - \\
  & FWHM (km s$^{-1}$) & 1944.9 $\pm$ 94.0 & - & 1571.0 $\pm$ 92.0 & 1857.1 $\pm$ 72.0 & 1708.7 $\pm$ 132.8 & 1658.0 $\pm$ 100.0 & - \\
  & Relative flux    & 0.52 $\pm$ 0.08 & - & 0.25 $\pm$ 0.02 & 1.00 & 0.15 $\pm$ 0.02 & 0.25 $\pm$ 0.03 & - \\
\hline
6 & Centroid (\AA)   & 4350.12 $\pm$ 0.79 & 4644.70 $\pm$ 1.88 & 4687.60 $\pm$ 1.30 & 4867.30 $\pm$ 0.30 & 5880.21 $\pm$ 1.19 & 6381.00 $\pm$ 2.00 & 6573.20 $\pm$ 0.20 \\
  & FWHM (km s$^{-1}$) & 2155.0 $\pm$ 127.0 & 2111.0 $\pm$ 310.0 & 1626.0 $\pm$ 208.0 & 2000.0 $\pm$ 50.0 & 1222.0 $\pm$ 143.0 & 820.0 $\pm$ 250.0 & 1541.0 $\pm$ 25.0 \\
  & Relative flux    & 0.44 $\pm$ 0.04 & 0.20 $\pm$ 0.04 & 0.20 $\pm$ 0.03 & 1.00 & 0.09 $\pm$ 0.02 & 0.03 $\pm$ 0.01  & 2.05 $\pm$ 0.09 \\
\hline
7 & Centroid (\AA)   & - & - & - & 4867.85 $\pm$ 0.65 & - & - & 6571.10 $\pm$ 0.50 \\
  & FWHM (km s$^{-1}$) & - & - & - & 1314.3 $\pm$ 95.0 & - & - & 1621.1 $\pm$ 67.0 \\
  & Relative flux    & - & - & - & 1.00 & - & - & 7.94 $\pm$ 0.95 \\
\hline
\end{tabular}
\end{sidewaystable*}
\end{appendix}


\end{document}